\documentclass{aa} 

\usepackage{graphicx}
\usepackage{txfonts}
\usepackage{lipsum}
\usepackage{subcaption}         
\usepackage{lscape}             
\usepackage{placeins}           
                                
\usepackage[dvipsnames]{xcolor}

\usepackage{hyperref}

\begin{document}

   \title{Long-term ultraviolet variability of Taurus T Tauri stars from the IUE archive}
   \subtitle{}

   \author{ Ana In\'es G\'omez de Castro\inst{1}$^,$\inst{2}\email{aig@ucm.es} }

   \institute{AEGORA Research Group -- Joint Center for Ultraviolet Astronomy, Universidad Complutense de Madrid, Plaza de Ciencias 3, 28040 Madrid, Spain
   \and Departamento de Fisica de la Tierra y Astrofisica, Fac. de CC. Matem\'aticas, Plaza de Ciencias 3, 28040 Madrid, Spain}
   \date{Received July 8, 2026}

  \abstract
{ Ultraviolet (UV) radiation from  Classical T Tauri stars is generated primarily by magnetospheric accretion and plays a fundamental role in the evolution of circumstellar disks. Despite its importance, the long-term variability of the UV radiation field remains poorly characterized because homogeneous time-series observations are scarce. We investigate the long-term variability of the near-UV continuum and the Mg\,II resonance doublet, and assess whether both diagnostics trace the same underlying accretion variability.}
{We analyzed the complete archive of long-wavelength International Ultraviolet Explorer (IUE) observations of T Tauri stars in the Taurus--Auriga star-forming region, comprising 475 low-dispersion spectra of 36 objects accumulated during the 18-year operational lifetime of the mission. The variability of the UV continuum was characterized using three continuum windows between 2600 and 3100~\AA. Continuum-subtracted Mg\,II line fluxes, UV--$V$ colors, and Lomb--Scargle periodograms were used to investigate the relationship between continuum and line variability and to search for characteristic timescales.}
{ The continuum fluxes measured within the 2675--3055~\AA\ interval are strongly correlated, with variability occurring
predominantly in continuum level rather than spectral shape. The near-UV and quasi-simultaneous optical fluxes are significantly
correlated in six of the seven best-sampled sources, although their relative variability differs among objects. The relation between
Mg~II and the near-UV continuum shows substantial object-to-object diversity, ranging from correlated variations with different
fractional responses to largely decoupled or anti-correlated behavior. Thus, Mg~II does not show a universal temporal response
to continuum variability. Lomb--Scargle periodograms show formally significant features in some of the best-sampled objects, including
features near 10 and 33~d. However, the irregular and clustered temporal sampling provides insufficient cycle coverage to establish
these features as periodic or characteristic timescales.}
{ }


\titlerunning{Accretion variability T Tauri stars}
\authorrunning{G\'omez de Castro}

\maketitle

\section{Introduction}\label{intro}

T Tauri stars (TTSs) are low-mass pre-main-sequence (PMS) stars characterized by strong magnetic activity and, in classical T Tauri
stars (CTTSs), ongoing accretion from a circumstellar disk. The UV--optical excess emission of CTTSs has long been recognized as a
signature of accretion. Early models attributed this excess to the release of accretion energy in a boundary layer between the disk and
the stellar surface \citep{1988ApJ...330..350B}, while optical monitoring revealed cool and hot spots whose rotational modulation
indicated an inhomogeneous stellar surface \citep[e.g.][]{1989A&A...211...99B}.

UV observations contributed importantly to the development of this picture. IUE monitoring of BP~Tau revealed variations of the UV
continuum and emission lines approximately in phase with the optical variability and required a hotter component than inferred from the
optical continuum alone \citep{1990AJ....100.1957S}. Shortly afterward, \citet{1991ApJ...370L..39K} formulated the  magnetospheric-accretion 
scenario, in which the stellar magnetic field truncates the inner disk and channels material toward the stellar surface. The resulting accretion 
shock produces high-energy radiation and an excess continuum extending from the UV into the optical \citep{1998ApJ...509..802C,1999MNRAS.304L..41G}. 
This excess is now widely used to estimate mass-accretion rates \citep{2013ApJ...767..112I,2021Natur.597...41E,2020A&A...639A..58M}.

Subsequent IUE monitoring of BP~Tau and DI~Cep revealed variability in both the UV continuum and emission lines, with different behaviors
among spectral tracers formed under different physical conditions \citep{1996MNRAS.283...55G,1997ApJ...482..465G}. These observations
showed that UV variability can probe rotational modulation and changes in the accretion flow. However, most previous UV variability studies
have focused on individual objects or relatively short observing sequences, leaving the long-term behavior comparatively poorly
constrained.

UV radiation is also important for protoplanetary-disk evolution, with the FUV radiation field contributing to disk photochemistry and
photoevaporation \citep{2017RSOS....470114E}. Disk irradiation may depend not only on the mean UV luminosity but also on its variability
\citep{2021MNRAS.508..950V,2024ApJ...970..118W}. The present study does not attempt to characterize the FUV or Ly$\alpha$ variability,
but focuses on the near-UV, where the IUE archive provides a sufficiently large set of repeated observations for a systematic
analysis. The Mg~II resonance doublet, which lies within this range, probes some of the same warm ($T\sim10^4$~K) chromospheric,
magnetospheric, and outflowing regions sampled by Ly$\alpha$, providing a complementary diagnostic of the T~Tauri environment.

The International Ultraviolet Explorer (IUE) operated for 18 years and provides a unique archive for investigating this variability.
By the end of the mission, Taurus--Auriga was the most extensively observed low-mass star-forming region in the IUE archive, with 475
low-dispersion ($\Delta\lambda\sim6$~\AA) spectra of 30 TTSs \citep{1997iueu.book.....G}. For comparison, HST has obtained UV
spectra of 29 Taurus TTSs, but repeated observations remain relatively scarce and were acquired with several instruments and observing modes
\citep{2025ApJ...985..109R}. The IUE archive therefore provides an exceptionally large set of UV spectra obtained with the same facility,
spanning timescales from individual observing campaigns to many years.

An additional advantage is that many IUE UV exposures were accompanied by optical measurements with the Fine Error Sensor (FES), generally
obtained immediately before the spectroscopic exposure. These measurements can be converted to an approximate $V$-band flux,
providing a large set of quasi-simultaneous UV--optical observations. This combination allows the long-term UV variability to be compared
directly with contemporaneous changes at optical wavelengths.

The primary goal of this work is to characterize the long-term near-UV variability of Taurus TTSs and its relation to the
quasi-simultaneous optical emission and to Mg~II variability. Section~2 describes the sample and IUE observations; Sects.~3--5
present the continuum, Mg~II, and time-series analyses, respectively; and Sect.~6 summarizes the conclusions.

\section{ IUE data sample of TTSs in Taurus-Auriga}

\begin{table*}
\caption{T Tauri stars in the Taurus-Auriga region observed with IUE in the LW range$^a$}.
\begin{tiny}
\begin{tabular}{lcccccccccc} 
\hline
 Name & Type$^b$ &\multicolumn{3}{c}{S.T.$^c$ }& V$^d$ & A$_V^e$ & Bin$^f$ &  NUV$^g$ &  F(Ly$_\alpha)^h$ & $\dot M_a ^i$\\
   &                   &          HBC &CDS& ULLYSES&         (mag.)& (mag.) &  & AB mag & $10^{-12}$ erg s$^{-1}$ cm$^{-2}$ & $M_{\odot}$ yr$^{-1}$  \\ \hline
{\bf AA Tau  }      & C &  K7 & K5Ve & K7            &12.37 &1.9  (0.74) & {\it S }             &            &$ 35.62 \pm 6.3$ & $1.5 \times 10^{-8}$\\ 
V773 Tau       & W &  K3 & K3Ve   &                  &10.62 &1.37$^{e1}$&{\it M}, 0."112  &            &                              & $2.9 \times 10^{-8}$\\
{\bf FM Tau}         & C &  M0 &M0/2e  & M0      &14.22 &2.7 (0.3)     & {\it M}, 37."3   & 17.88  &                              & $1.2 \times 10^{-9}$ \\
{\bf CW Tau}         & C &  K3 &K0Ve  & K0       &12.36 &2.6              & {\it S}              & 17.92 &                               & $3.4 \times 10^{-8}$\\
{\bf V410 Tau}       & W &  K3 & K3Ve   &         &10.82 &0.66             & {\it D}, 0."123 &            &                               &                               \\
V1069 Tau & W &  K7e &    &                          &12.28 &0.00$^{e2}$  & {\it S}             &             &                              &                              \\
{\bf BP Tau }        & C &  K7 &K5/7e  & K7       &12.09 &1.37 (0.5)     & {\it S }            & 15.97 & $31.5 \pm 5.5$       & \\
{\bf DE Tau  }       & C &  M2 &M3Ve& M2        &13.00 &1.62   (0.6)   & {\it S}             & 16.49 & $ 15.4 \pm 4.5 $       & $1.2 \times 10^{-9}$ \\
{\bf RY Tau  }    & C &  K1 &K1IV/Ve & G1        &10.01 &2.7  (1.8)     & {\it S}             & 15.38 &                                & $7.76 \times 10^{-8}$\\
HD 283572       & {\bf W} &  G5 &  G5IVe  &             &9.04 &0.3$^{e3}$     &{\it  S}             &15.63    &                               &\\
{\bf T Tau  }        & C &  K0 &  K0IV/Ve  &          &9.90 &2.1   (1.5)      & {\it D}, 0."71  & 14.97 & &\\
{\bf DF Tau  }     & C  &M0,1& M3Ve& M2         &11.49 &1.96  (0.6)    & {\it D}, 0."088 & 16.78 & & $1.2 \times 10^{-9}$\\
{\bf DG Tau  }       & C &  M  &K6Ve  & K7        &12.01 &1.6                & {\it S}             & & & $3.5 \times 10^{-8}$\\
HD 285778& W &   &   K1e &                          &10.34 &          (0.0)   & {\it S}              &   & &\\
{\bf UX Tau  }    & {\bf C} &  K2 &  K2Ve  & K2        &10.69 &1.8               & {\it   M}, 5"         & & $6.2 \pm 0.5$  & $4.73 \times 10^{-8}$\\
{\bf DK Tau  }  & C &  K7 &K7e+K7e& K5        &11.70 &2.01              & {\it D}, 2."53    & & & $3.4 \times 10^{-8}$\\
V1075 Tau& W &  K7 &  K7e  &                      &12.06 &0.00$^{e2}$   & {\it S}              & & &  \\
{\bf V827 Tau } &W&K7,M0&K7Ve&  K5         &12.18 &1.24 (1.1)      & {\it S}              & & &   \\
{\bf V826 Tau } &W& K7,M0&K7.0Ve& K7      &12.11 &1.05 (0.3)      & {\it SB}, 0."9E-4 & & &  \\
{\bf UZ Tau  A} &C&M1,3&M4Ve+M1/3Ve&M1&12.86 &2.86              & {\it M}, 3."78.   &  & & $9.5 \times 10^{-8}$\\
GG Tau         & C &  K7 &M0e+M2.0e  &.        & 12.24 &1.94             & {\it M}, 10."3    & & &  \\
GH Tau         & C &  M2,3&  M3Ve &                &12.95 &1.06              & {\it D}, 0."314  & 18.25 && \\
V807 Tau       & C &  K7 & K7   &                    &11.22 &0.65$^{e2}$.  & {\it M}, 0."41    & 17.76& &  \\
{\bf V830 Tau  }  & W &  K7,M0&M0-1& M0-1 &12.21 &0.77 (0.7)     & {\it S}               & & &   \\
GI Tau         & C &  K6 & K5Ve &                     &13.01 &2.52  (1.3)     & {\it M}              & & &  \\
GK Tau         & C &  K7 &K7Ve  &                    &12.14 &1.8 (0.9)       & {\it M }              & & &  \\
DL Tau         & C &  K7 & K7Ve &                     &13.05 &1.53 (1.3)     & {\it S}               & & &  \\
{\bf DN Tau  }       & C &  M0 & M1:Ve & M0    &12.36 &1.19  (1.9)    &{\it  S }              & & & $1.0 \times 10^{-8}$\\
Haro 6-37      & C &  K6 &   K7e+M1e &           &13.55 &1.8$^{e4}$    & {\it D}, 2."7       & & &\\
{\bf DR Tau    }     & C &Cont.  &K5Ve& K5      &11.17 &1.6  (1.2)       & {\it S}               & 14.87 & & $5.2 \times 10^{-8}$ \\
{\bf DS Tau   }      & C &  K5 & K4Ve & K5       &11.90 &1.43 (0.3)      & {\it S}                & & & $2.7 \times 10^{-9}$ \\
UY Aur         & C &  K7 & M0e+M2.5e &          &12.37 &1.88              & {\it D}, 0."89     & 16.29 & & \\
{\bf GM Aur }        & C &  K3 & K3Ve & K7       &12.03 &0.5   (0.3)     & {\it S} & 16.39 & $12.2 \pm 3.0$ & $7.0 \times 10^{-9}$\\
{\bf LkCa 19  }      & W &  K0 &  K0Ve  & K0.  & 10.85 &0.65$^{e2}$ & {\it S}               & & & $1.7 \times 10^{-8}$\\
{\bf SU Aur   }      & {\bf C} &  G2 &   G2IIIne &  G5 & 8.93 &1.45  (0.9)     & {\it S }              & 15.21 &$35.4 \pm 8.2$ &  \\
{\bf RW Aur   }  &  C &  K1 & K1/5e+K5e & K0&10.12 &     (1.2)       & {\it M}, 1."50    & 14.13 & & $2.0 \times 10^{-8}$\\ \hline
\end{tabular}

Tablenotes: {\bf $(a)$: } Stars included in the Hubble UV Legacy Library of Young Stars as Essential Standards (ULLYSES) \url{https://ullyses.stsci.edu/index.html} sample are marked in bold. {\bf $(b)$:} Type stands for: W$\equiv$ WTTS, C$\equiv$CTTS.  {\bf $(c)$: }
Spectral types from the  Herbig and Bell Catalogue (HBC) \citep{1988cels.book.....H},  the Centre de Donn\'ees Stellaires (CDS) \url{https://simbad.cds.unistra.fr/simbad/sim-fid}.
and  ULLYSES.{ \bf $(d)$ :} V from the HBC. { \bf $(e)$: } baseline compilations are \citet{1990AJ.....99..924B} and \citet{2012ApJ...744..121Y} - 
for the rest: e1 \citep{1995AJ....110..776W}, e2 \citep{1995ApJ...446..331D}, e3 \citep{1987ApJ...314..297W}, and e4 \citep{1988AJ.....95..534S}.
{\bf $(f)$: } {\it S} stands for single, {\it SB} stands for spectroscopic binary (the a$\sin i$ value is given in arcsec), {\it D} stands for  double 
(the distance to the companion is provided),  {\it M} stands for multiple system (the distance  to the farthest component is given). Source: \citet{1997iueu.book.....G}.
{\bf $(g)$: } GALEX NUV magnitude from \citet{2015ApJS..216...26G}.  {\bf $(h)$: } Ly$\alpha$ flux  from \citet{2012ApJ...756L..23S}. {\bf $(i)$: } accretion rates from ULLYSES.
\end{tiny}
\end{table*}

A total of 42 TTSs were observed with the IUE  in the long-wavelength range (2000--3200 \AA) during its operational lifetime. However, the spectra of six stars---V819 Tau (HBC 378), DI Tau (HBC 39), HBC 392, HL Tau (HBC 49), HN Tau (HBC 60), and DQ Tau (HBC 72)---are severely underexposed and therefore unsuitable for quantitative analysis. The working sample is thus reduced to 36 TTSs: 10 weak-lined TTSs (WTTSs), 26 classical TTSs (CTTSs). 

The temporal coverage of the WTTSs is considerably more limited than that of the CTTSs. WTTSs are generally weaker UV sources and most
have only a few usable IUE spectra. Their variability therefore cannot be characterized with the same reliability as that of the
better-sampled CTTSs, precluding a meaningful statistical comparison between the two classes.

All available IUE low-dispersion spectra obtained with the LWP and LWR cameras (475 spectra in total) have been retrieved. A complete list of spectra, including observation dates, exposure times, and additional metadata, is given by \citet{1997iueu.book.....G}. Table~1 summarizes the stellar properties most relevant to the present analysis. Because the IUE large aperture is 10"$\times$20" and the spatial resolution along the slit is $3"$, the UV spectra of binary and some multiple systems typically correspond to the combined emission of the system, generally dominated by its brightest UV component. 

The spectra were retrieved from the IUE Final Archive and processed with the New IUE Spectral Image Processing System (NEWSIPS), the standard reduction pipeline for the final archive. However, NEWSIPS is known to underestimate emission-line fluxes in spectra with weak continua. Therefore, for objects displaying faint continua but strong Mg II emission, we adopted spectra reduced with the IUE Newly Extracted Spectra (INES) system, which provides a more reliable extraction of emission-line fluxes in these cases (\citealt{1998ESASP.413..735S}). A detailed comparison between the NEWSIPS and INES reductions is presented by \citet{2000MNRAS.312..833H}. No extinction correction was applied because extinction toward TTSs is highly variable and quasi-simultaneous determinations of $A_V$ are generally unavailable. Moreover, the extinction law of circumstellar material around TTSs remains poorly constrained, introducing additional uncertainties into any correction.

\begin{figure*}
\centering
 \includegraphics[width=20cm]{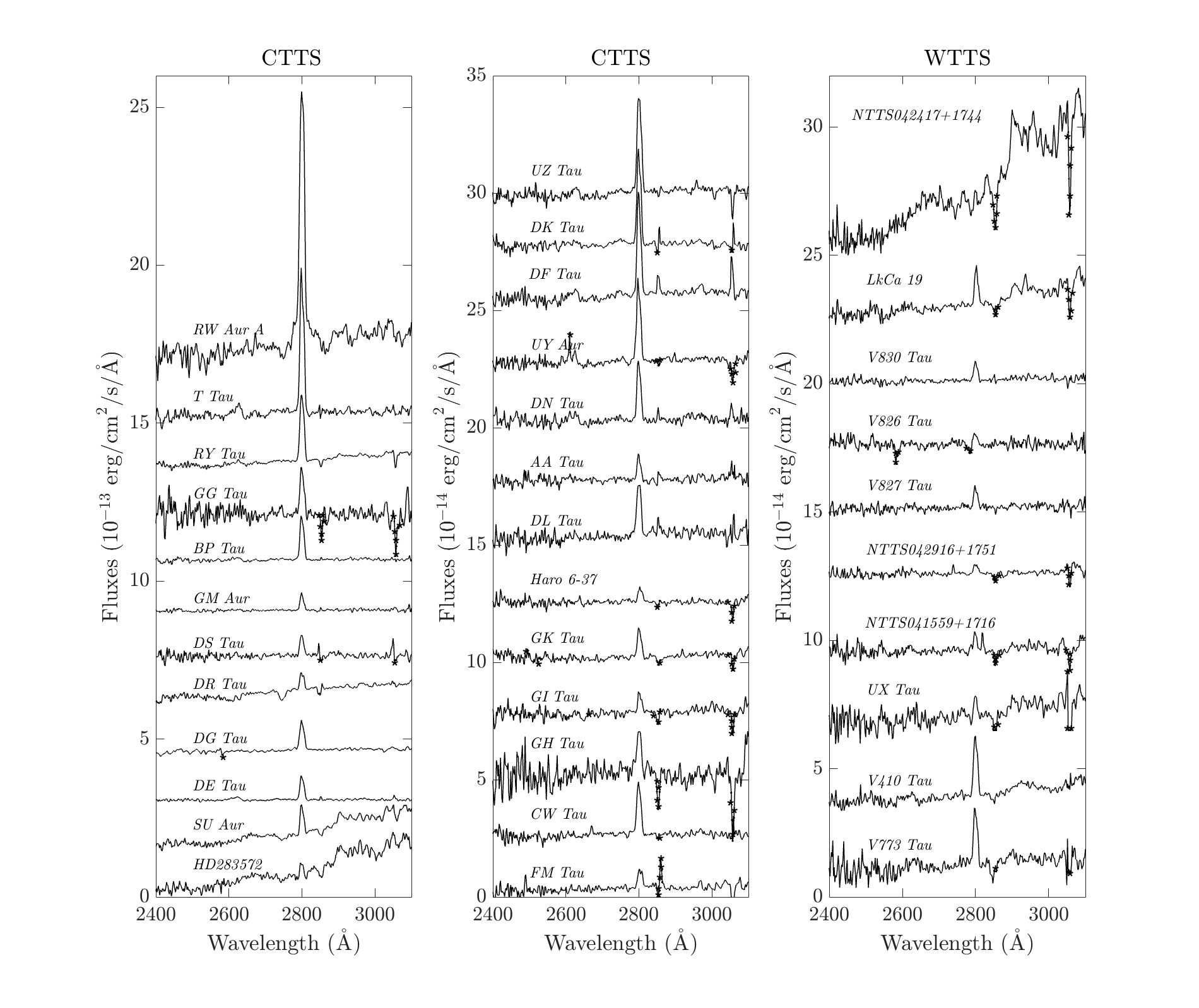} 
   \caption{Average UV spectrum of the TTSs observed
with the IUE in the Taurus star forming region. The strong Mg II
emission feature is observed in most of the sources. The asterisks mark the
location of bad pixels. Continuum offsets have been applied to the spectra for display purposes.}
\end{figure*}

Figure~1 shows the mean UV spectrum of each TTS in the sample in the 2600--3100~\AA\ wavelength interval. This region was selected because it provides the highest signal-to-noise ratio in the IUE long-wavelength spectra while avoiding the rapid decline in instrumental sensitivity at shorter wavelengths. The mean spectrum was computed by averaging all unsaturated observations, with bad pixels identified by asterisks. It is characterized by a weak continuum together with several prominent spectral features, including the Mg II resonance doublet at 2800~\AA, Mg I at 2852~\AA, and Fe II (UV1--3) (see, e.g., Fig.~1 in \citealt{2026SSRv..222...52G}). The Mg II emission is detected in nearly all objects and is frequently strong enough to saturate the spectra of classical TTSs. In contrast, the Fe II multiplets and, particularly, the C II] emission are detected in only a few stars and display substantial variability.

\section {The UV continuum}

For this analysis, the sample is reduced to 30 stars, since six of the objects listed in Table~1 (Haro 6-37, V807 Tau, GH Tau, V826 Tau, GG Tau, and NTTS 041559+1716) exhibit underexposed continua. Multiple windows were used to determine whether variations in the continuum flux are accompanied by changes in its spectral shape, as expected from accretion-shock models \citep[e.g.][]{2013ApJ...767..112I,2018MNRAS.475.4367D,2024MNRAS.531.3744K}.
The windows are 50~\AA\ wide and centered at 2700, 2900, and 3030~\AA. They lie shortward of the Balmer edge, where the continuum
shape is sensitive to the relative contributions of the heated photosphere and Balmer continuum, and avoid regions contaminated by
strong Fe~II, Mg~II, Mg~I, or molecular absorption \citep{2026ApJ..1009L..11G}. Only spectra with SNR $>3$ in the corresponding window were retained.

The UV continuum varies by factors of $\sim2$--10. Figure~2 shows the 2700 and 3030~\AA\ continuum fluxes against the 2900~\AA\ flux for
RY~Tau, the most frequently observed star with IUE; corresponding plots for the other sources are shown in Appendix~A. Despite variations
of almost a factor of two, the RY~Tau continuum fluxes are strongly correlated. Least-squares fits yield slopes of $1.67\pm0.08$ for
$F_{2900}/F_{2700}$ and $1.17\pm0.05$ for $F_{3030}/F_{2900}$, with small intercepts relative to the observed flux range. 
These UV flux ratios are relatively insensitive to extinction: adopting the Galactic extinction law of \citet{2023ApJ...950...86G} with $R_V=3.1$ gives
$A_\lambda/A_V=2.05$, 1.90, and 1.83 at 2700, 2900, and 3030~\AA, respectively, corresponding to a change in
$\log(F_{270}/F_{303})$ of only $\simeq0.09 A_V$ dex.

The same analysis for the other well-sampled TTSs ($N>10$; BP~Tau, RW~Aur, DR~Tau, T~Tau, DF~Tau, and SU~Aur) yields strong
linear correlations with $p\ll0.01$ (Table~2). Thus, within the wavelength interval sampled here, continuum variations are dominated
by changes in amplitude rather than by large changes in spectral shape. The characteristic continuum shape nevertheless differs among
sources, as shown by the broad range of mean flux ratios in Fig.~3.

BP~Tau provides an interesting long-term comparison. Recent HST/ULLYSES observations \citep{2025ApJ...985..109R} show similarly
strong correlations among the continuum bands, but different mean flux ratios from those measured with IUE (Fig.~3), suggesting a change
in its characteristic near-UV continuum shape between the two observing epochs.

We therefore use a single continuum window as a proxy for the continuum level over the wavelength interval sampled here.

\begin{figure}
\centering
   \includegraphics[width=6cm]{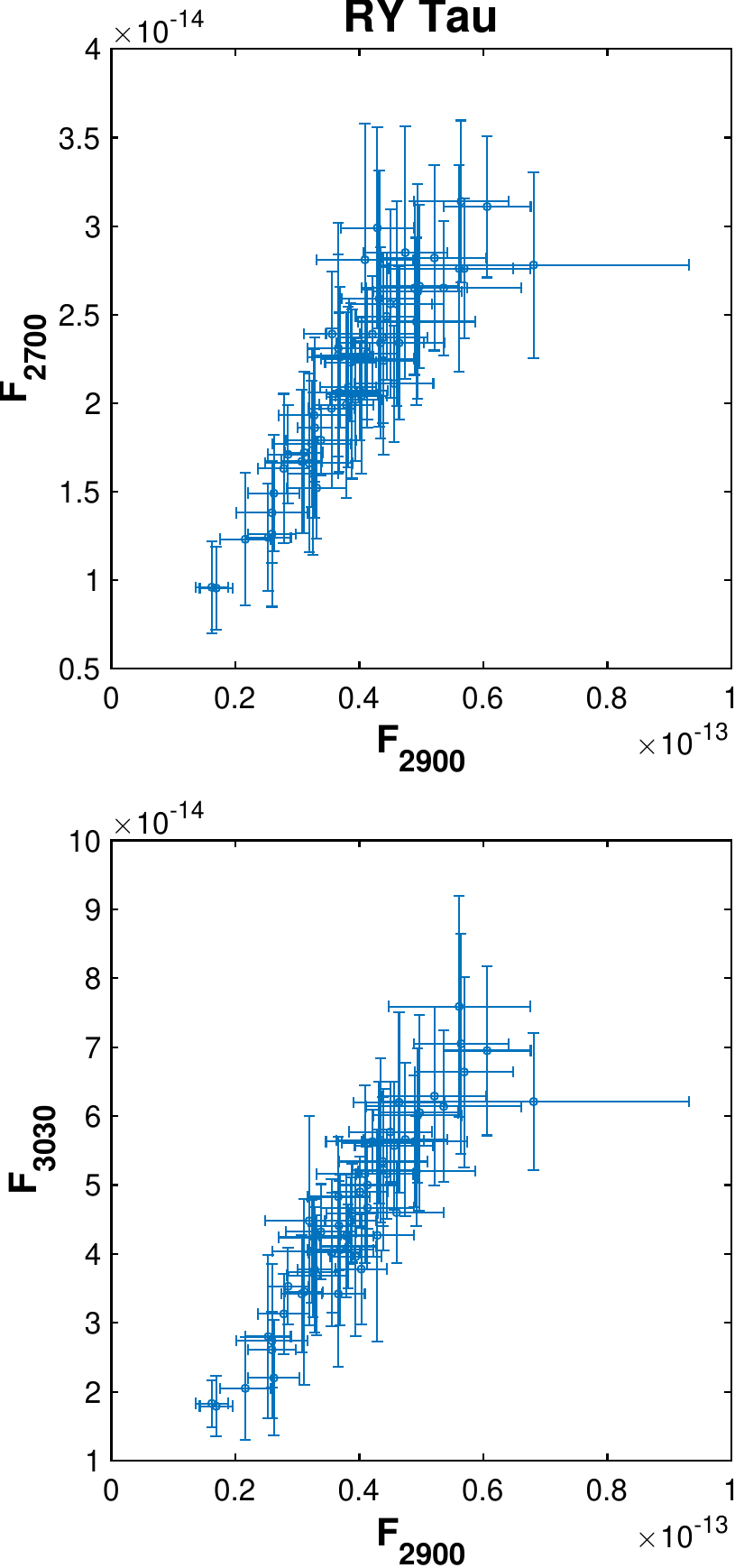}\\
   \caption{Variation of the UV continuum energy distribution for  all the observations of RY Tau with a well exposed continuum.
{\it Top:} $F_{2700}$ vs $F_{2900}$ flux and {\it bottom:} $F_{3030}$ vs $F_{2900}$ flux. The fluxes are given in units of 10$^{-14}$ erg cm$^{-2}$ s$^{-1}$ \AA$^{-1}$.
}
\end{figure}

\begin{figure}
\centering
   \includegraphics[width=11 cm]{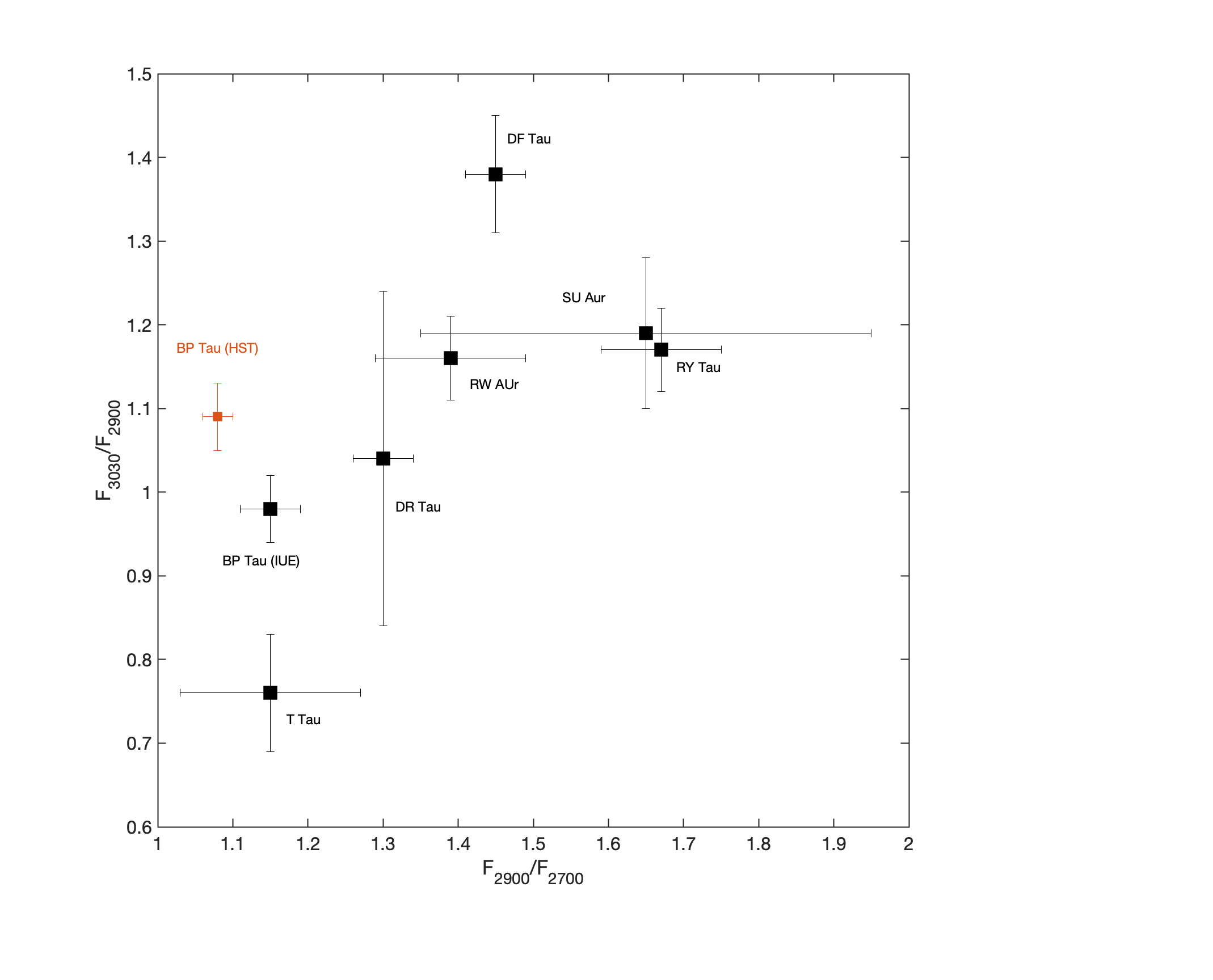}
   \caption{Comparison of the continuum scaling relations listed in Table 2. Each point corresponds to one star. The horizontal axis shows the slope of the
   $F_{2900}/F_{2700}$  relation, while the vertical axis shows the slope of the $F_{3030}/F_{2900}$ relation. }
\end{figure}

\begin{table*}

\caption{Variability of the UV Continuum}
\begin{center}
\begin{tabular}{l|ccc|ccc|c}
\hline
Name &\multicolumn{3}{c}{$F_{2900}$  vs.  $F_{2700}$ }& \multicolumn{3}{c}{$F_{3030}$ vs. $F_{2900}$} &N$^a$\\

     & Slope   & Intercept   & Pearson R  & Slope   & Intercept   & Pearson R    \\
     &     & { \tiny $10^{-14}$erg~cm$^{-2}$~s$^{-1}$~\AA\ $^{-1}$ } & & & 
{ \tiny $10^{-14}$erg~cm$^{-2}$~s$^{-1}$~\AA\ $^{-1}$ }& &\\
\hline

BP Tau (b)& 1.15$\pm$ 0.04 & 0.04$\pm$0.06  & 0.973 & 0.98$\pm$0.04 & 0.04$\pm$0.07
& 0.987 & 53 \\
 &{\tiny $(1.08 \pm  0.02)$} &  {\tiny ($0.15 \pm 0.06$ )} & {\tiny (0.994)} & {\tiny  ($1.09 \pm 0.04 $) } & {\tiny ($0.08 \pm 0.10$)} & {\tiny (0.999)} & {\tiny (26)} \\
RW Aur & 1.39$\pm$ 0.1  & 0.20$\pm$1.16  & 0.945 & 1.16$\pm$0.05 &-0.72$\pm$0.73
& 0.984 & 23 \\
DR Tau & 1.30$\pm$ 0.04 & 0.25$\pm$0.18  & 0.988 & 1.04$\pm$0.05 & 0.20$\pm$0.32
& 0.967 & 35 \\
RY Tau & 1.67$\pm$ 0.08 & 0.35$\pm$0.16  & 0.932 & 1.17$\pm$0.05 &-0.08$\pm$0.20
& 0.936 & 60 \\
DF Tau & 1.45$\pm$ 0.04 &-0.04$\pm$0.04  & 0.995 & 1.38$\pm$0.07 &-0.29$\pm$0.10
& 0.982 & 10 \\
SU Aur & 1.65$\pm$ 0.30 & 1.99$\pm$1.21  & 0.795& 1.19$\pm$0.09 & 0.52$\pm$0.79
& 0.952 & 20 \\
T Tau  & 1.15$\pm$ 0.12 & 0.76$\pm$0.34  & 0.913 & 0.76$\pm$0.07 & 0.80$\pm$0.28
& 0.931 & 10 \\ 
 \hline
\end{tabular}
\begin{tabular}{l}
{\tiny (a) Number of spectra used; (b) The scalings vary slightly during the ULLYSES observations of BP Tau (see values into brackets).} \\
\end{tabular}
\end{center}
\end{table*}

\vspace{0.5cm}
\noindent
{\it Photospheric contribution: the UV-V color}
\vspace{0.2cm}

Since the photospheric contribution is significantly larger in the optical than in the UV for cool TTSs, quasi-simultaneous UV and optical
measurements provide information on changes in the relative continuum emission at the two wavelengths. We use the optical measurements
obtained with the IUE Fine Error Sensor (FES), which recorded a visual measurement immediately before each UV exposure. The FES magnitudes, stored in the image headers, have rms uncertainties of 0.08~mag and a reproducibility of about 0.04~mag \citep{1981IUENN..15...74H}.
Although the FES bandpass is similar to the Johnson $V$ band, its magnitudes are not directly on the standard $V$ system.

     The FES count rates were converted into equivalent $V$ magnitudes following the calibration of \citet{1991EIUEN..38...27P}, including corrections for color terms, the long-term sensitivity degradation of the cameras, and the changes in the FES reference point introduced after January 1990 at GSFC and July 1990 at VILSPA. The corresponding optical fluxes were computed as
$$
 F_{\rm FES} = F_{5556}({\rm Vega}) 10^{-0.4 V_*}
$$

Figure~4 compares the near-UV continuum flux, $F_{2900}$, with the quasi-simultaneous optical flux, $F_V$, derived from the FES magnitude for
the seven sources with at least ten valid measurements. A positive trend is apparent in most sources, although the strength and scatter
of the relation vary considerably from object to object.

Unlike the comparison among the three near-UV continuum bands, we do not fit a linear relation between $F_{2900}$ and $F_V$. The
UV--optical flux--flux diagrams show substantially larger scatter, making the slope of such a fit poorly constrained and of limited
physical significance. Moreover, the optical and near-UV fluxes contain different relative contributions from the stellar photosphere
and accretion emission, so there is no reason to expect them to scale proportionally. We therefore characterize their degree of association
using the Pearson ($r_{\rm P}$) and Spearman ($\rho_{\rm S}$) correlation coefficients and their two-sided probabilities.
A Benjamini--Hochberg false-discovery-rate (FDR) correction was applied to the seven tests; the results are
summarized in Table~\ref{tab:UVVcorr}. Significant positive correlations ($p_{\rm FDR}<0.05$) are found with both tests in six of
the seven sources. The Pearson coefficients range from 0.54 for BP~Tau to 0.95 for DR~Tau and RW~Aur. DF~Tau formally gives 
$r_{\rm P}=0.98$, although one of its ten measurements has substantial leverage; its high Spearman coefficient ($\rho_{\rm S}=0.88$)
nevertheless supports a positive monotonic relation. T~Tau is the only source without a statistically significant correlation
($r_{\rm P}=0.53$, $p=0.12$; $\rho_{\rm S}=0.49$, $p=0.15$).

The prevalence of positive correlations indicates that the near-UV and optical variability are coupled in most of the best-sampled
sources. The optical flux, however, should not be regarded as purely photospheric, since accretion-shock emission can contribute
significantly at $V$-band wavelengths \citep{2025ApJ...993..181P}. Moreover, the relative accretion contribution near 2900 and
5500~\AA\ depends on the properties of the accretion columns \citep{2013ApJ...767..112I}, providing a possible origin for the
different degrees of UV--optical coupling among sources.

To characterize changes in the relative amplitudes of the UV and optical emission, we define
\begin{equation}
\langle {\rm UV-V} \rangle =
\frac{1}{N}\sum_{i=1}^{N}
\left[-2.5\log
\left(\frac{F_{2900,i}}{F_{{\rm FES},i}}\right)\right].
\end{equation}
This quantity is analogous to a color index but is derived from quasi-simultaneous near-UV and optical measurements. Its temporal dispersion,
computed as the sample standard deviation, measures the observed long-term variability of the near-UV emission relative to the optical
continuum; no correction for individual measurement uncertainties was applied. Sources with only one or two valid observations are flagged
in Table~4.

The mean fluxes, UV--V indices, and dispersions are summarized in Table~4. Together with the correlation analysis, they show that the
near-UV and optical fluxes generally vary together in the best-sampled sources, but not necessarily in proportion. This behavior is
consistent with variable accretion emission contributing at both near-UV and optical wavelengths, with its relative contribution and
spectral energy distribution varying among sources and epochs.

Chromospheric variability may also contribute to the observed UV variations. In practice, the distinction between accretion- and
activity-dominated emission does not translate into a sharp observational boundary between CTTSs and WTTSs, particularly at low
accretion levels \citep{2012ApJ...749..190G}. Unfortunately, the low S/N and sparse IUE coverage of the WTTSs in our sample preclude a
statistical assessment of the possible contribution from magnetic activity.

\begin{table}
\caption{Correlation between the near-UV and optical fluxes for
sources with at least ten quasi-simultaneous measurements.}
\label{tab:UVVcorr}
\centering
\small
\begin{tabular}{lccccc}
\hline\hline
Star & $N$ & $r_{\rm P}$ & $p_{\rm P,FDR}$ &
$\rho_{\rm S}$ & $p_{\rm S,FDR}$ \\
\hline
BP Tau & 38 & 0.543 & $4.98\times10^{-4}$ & 0.442 & $6.33\times10^{-3}$ \\
DF Tau & 10 & 0.978 & $6.75\times10^{-6}$ & 0.875 & $2.31\times10^{-3}$ \\
DR Tau & 35 & 0.950 & $2.02\times10^{-17}$ & 0.944 & $7.08\times10^{-17}$ \\
RW Aur & 18 & 0.953 & $1.12\times10^{-8}$ & 0.940 & $6.37\times10^{-8}$ \\
RY Tau & 60 & 0.791 & $2.25\times10^{-13}$ & 0.753 & $1.34\times10^{-11}$ \\
SU Aur & 20 & 0.751 & $3.61\times10^{-4}$ & 0.746 & $5.07\times10^{-4}$ \\
T Tau  & 10 & 0.528 & 0.117 & 0.491 & 0.154 \\
\hline
\end{tabular}

\tablefoot{
$r_{\rm P}$ and $\rho_{\rm S}$ are the Pearson and Spearman
correlation coefficients, respectively. The associated probabilities
are two-sided and have been corrected for multiple comparisons using
the Benjamini--Hochberg false-discovery-rate (FDR) procedure.
}
\end{table}

\begin{table}
\caption{The UV-V color}
\small
\begin{tabular}{lrrrrrrr}
\hline
\\
Name  & N & $<$F$_{2900}>$ &$<$ F$_V>$ & <UV-V> \\
             &   &  \multicolumn{2}{c}{$10^{-14} \frac{erg}{cm^2s\AA}$} & \\ 
             \\ \hline
             
AA Tau$^{a}$   &  1 & 0.51  $\pm$ 0.14  &  3.27 $\pm$ 0.12  & 2.02  $\pm$ 0.30  \\
BP Tau         & 38 & 1.87  $\pm$ 0.71  &  6.4  $\pm$ 1.1   & 1.39  $\pm$ 0.33  \\
DE Tau         &  6 & 0.763 $\pm$ 0.096 &  3.22 $\pm$ 0.18  & 1.57  $\pm$ 0.10  \\
DF Tau         & 10 & 1.2   $\pm$ 1.0   &  6.4  $\pm$ 2.4   & 1.97  $\pm$ 0.40  \\
DG Tau         &  7 & 1.74  $\pm$ 0.33  &  5.5  $\pm$ 1.0   & 1.256 $\pm$ 0.086 \\
DK Tau$^{a}$   &  1 & 0.66  $\pm$ 0.13  &  4.15 $\pm$ 0.15  & 1.99  $\pm$ 0.22  \\
DL Tau$^{a}$   &  1 & 0.802 $\pm$ 0.070 &  3.35 $\pm$ 0.12  & 1.55  $\pm$ 0.10  \\
DN Tau$^{a}$   &  2 & 0.40  $\pm$ 0.13  &  4.937 $\pm$ 0.042 & 2.76 $\pm$ 0.37  \\
DR Tau         & 35 & 6.4   $\pm$ 2.4   & 11.0  $\pm$ 3.2   & 0.63  $\pm$ 0.17  \\
DS Tau         &  3 & 1.9   $\pm$ 1.0   &  6.4  $\pm$ 2.3   & 1.38  $\pm$ 0.19  \\
GM Aur$^{a}$   &  2 & 0.56  $\pm$ 0.18  &  5.52 $\pm$ 0.25  & 2.51  $\pm$ 0.40  \\
HD 283572      &  5 & 11.1  $\pm$ 3.8   & 73    $\pm$ 31    & 1.98  $\pm$ 0.22  \\
RW Aur         & 18 & 14.3  $\pm$ 7.6   & 30    $\pm$ 14    & 0.89  $\pm$ 0.31  \\
RY Tau         & 60 & 3.9   $\pm$ 1.1   & 43    $\pm$ 11    & 2.60  $\pm$ 0.20  \\
SU Aur         & 20 & 8.6   $\pm$ 1.3   & 76.3  $\pm$ 6.4   & 2.37  $\pm$ 0.12  \\
T Tau          & 10 & 4.2   $\pm$ 1.7   & 40.9  $\pm$ 3.0   & 2.55  $\pm$ 0.40  \\
UX Tau$^{a}$   &  1 & 1.19  $\pm$ 0.30  & 19.79 $\pm$ 0.73  & 3.05  $\pm$ 0.28  \\
V410 Tau$^{a}$ &  2 & 0.727 $\pm$ 0.014 & 17.6  $\pm$ 1.4   & 3.457 $\pm$ 0.067 \\

\hline
\end{tabular}
\tablefoot{
$^{a}$ For sources with only one or two valid observations, the quoted
uncertainties represent the rms uncertainties of the individual measurements
rather than the intrinsic temporal dispersion.
}
\end{table}

\begin{figure*}
\includegraphics[width=18cm]{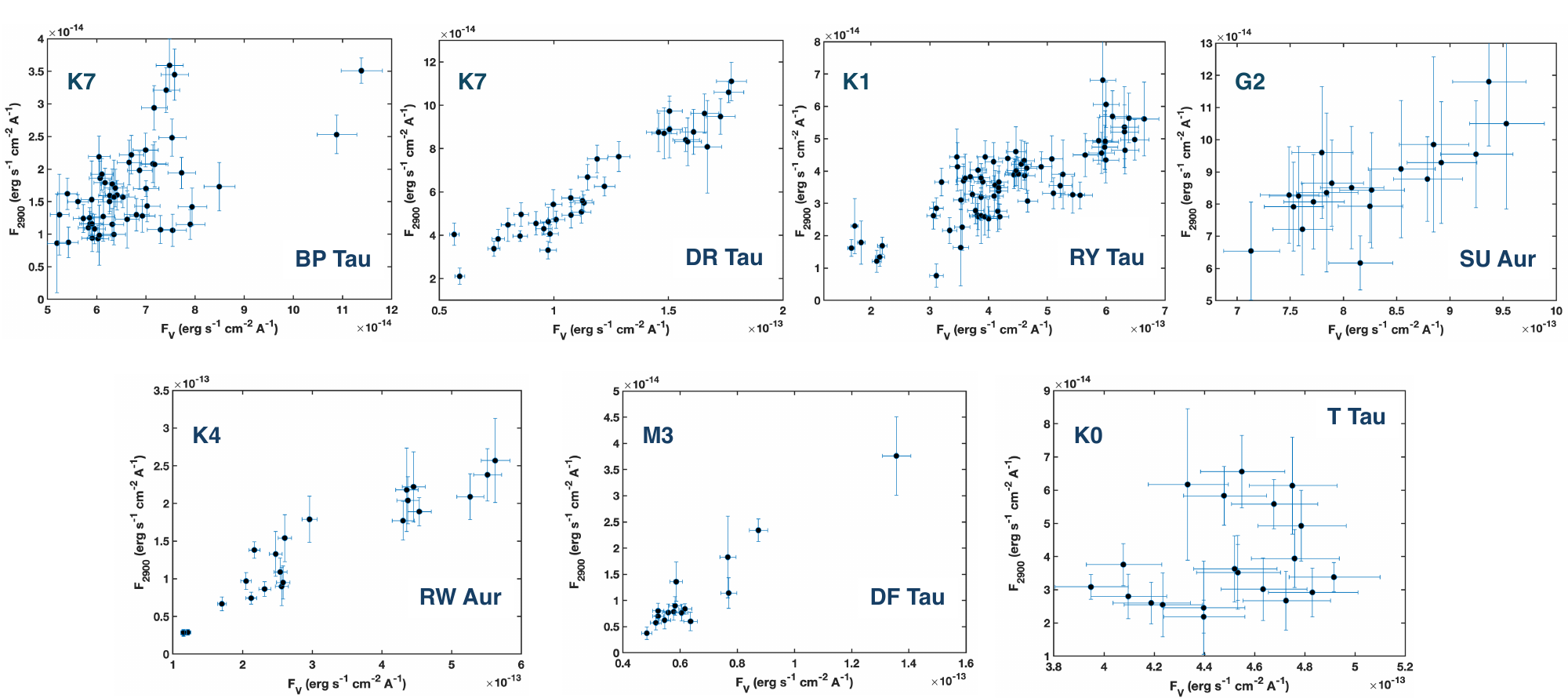}
   \caption{ Flux--flux diagram comparing the continuum flux at 2900~\AA\ with the quasi-simultaneous optical flux derived from the IUE Fine Error
Sensor. A positive relation is apparent in most of the best-sampled sources, although its strength and scatter vary substantially from
object to object. T~Tau is the only source in this subset for which the correlation is not statistically significant (see
Table~\ref{tab:UVVcorr}).}
\end{figure*}

\section{The Mg II emission}

The Mg~II resonance doublet is one of the strongest UV emission features in TTSs and has long been used as a diagnostic of
magnetospheric accretion and outflow activity \citep{1981ApJ...251..113G,2002ApJ...567.1013A,2015MNRAS.448..484L}.
Its emission may arise from several physically distinct regions, including the chromosphere and extended magnetospheric plasma, while 
the observed profiles can be strongly affected by winds and outflows \citep{2015MNRAS.448..484L}. Spatially resolved HST observations also
show that TTS jets can be strong Mg~II emitters \citep{2007ApJ...663..350C,2012ApJ...749..139C}. Thus, unresolved IUE
spectra may contain contributions from chromospheric, magnetospheric, and outflowing plasma that need not vary with the same amplitudes or
timescales. We therefore compare the Mg~II emission with the near-UV continuum to investigate whether the two diagnostics vary together and to what
extent they may trace the same underlying physical processes.

For each IUE spectrum with unsaturated Mg~II emission, the local continuum was estimated by fitting a linear function to two
line-free windows (2780--2790 and 2810--2820~\AA). This fit was subtracted over the Mg~II integration interval
(2792--2805~\AA), and the residual emission was integrated to obtain the net Mg~II flux.

For each observation, we derived: {\it (i)} the UV continuum color, $F_{2700}/F_{3030}$; {\it (ii)} the net Mg~II flux; and {\it (iii)}
the ratio of the net Mg~II flux to the underlying continuum at line center. For each star we computed the mean, median, standard
deviation, and median absolute deviation (MAD) of these quantities, together with the number of spectra, observing baseline, and median
cadence. The MAD provides a robust measure of the dispersion that is less sensitive to isolated measurements.

To quantify the intrinsic Mg~II variability while accounting for measurement uncertainties, we computed the fractional variability
amplitude,
\begin{equation}
F_{\rm var} =
\frac{\left(\sigma_{\rm obs}^2-\langle\epsilon^2\rangle\right)^{1/2}}
{\langle F\rangle},
\end{equation}
where $\sigma_{\rm obs}$ is the observed standard deviation, $\langle F\rangle$ the mean Mg~II flux, and
\begin{equation}
\langle\epsilon^2\rangle =
\frac{1}{N}\sum_{i=1}^{N}\epsilon_i^2
\end{equation}
is the mean squared measurement uncertainty. When $\sigma_{\rm obs}^2\leq\langle\epsilon^2\rangle$, $F_{\rm var}$ is
set to zero. Figure~5 provides an overview of the resulting variability amplitudes and temporal sampling. We adopt
approximately ten observations as a pragmatic threshold for identifying the better-sampled sources. This is not a formal
statistical limit, but reduces the sensitivity of $F_{\rm var}$ to individual measurements. About ten sources satisfy this criterion,
although the number of observations alone does not describe the temporal coverage because IUE observations are often clustered into
campaigns separated by long gaps.

The relation between the mean Mg~II emission and near-UV continuum across the sample was investigated in logarithmic space assuming
\begin{equation}
F({\rm MgII})=A\,F_{\rm cont}^{\alpha}.
\end{equation}
The regression was performed using weighted least squares, with the uncertainty in each mean flux estimated as the MAD divided by the
square root of the number of observations. The resulting ensemble relation,
\begin{equation}
\alpha=0.82\pm0.14,
\end{equation}
is shown in Fig.~6. The sub-linear relation indicates that the mean Mg~II flux increases less rapidly than the mean near-UV continuum
across the sample. This behavior may reflect the optical thickness of the Mg~II resonance lines and/or differences in the relative
contributions of the Mg~II-emitting components among sources.  The slope is also comparable to the range of relations between
Mg~II and accretion luminosity derived from accretion-shock modeling \citep{2013ApJ...767..112I}. However, this population-level relation
does not imply a similar temporal response within individual objects.

Figure~7 and Table~5 show the corresponding relations within individual stars, fitted as $\log F_{\rm MgII}=\alpha\log F_{\rm cont}+\beta$, together with the
Pearson and Spearman correlation coefficients and the rms residuals.
RY~Tau and RW~Aur show the clearest correlated variability, with $r_{\rm P}=0.91$ and 0.83 and slopes of $\alpha=0.57\pm0.03$ and $1.39\pm0.23$, respectively. The RY~Tau slope is sensitive to a single high-flux observation outside the otherwise limited continuum range, so the individual slopes are best
regarded as empirical descriptions of the observed variability.

More generally, the fitted $\alpha$ values span a broad range. Positive values indicate correlated variations, with
$0<\alpha<1$, $\alpha\simeq1$, and $\alpha>1$ corresponding, respectively, to weaker, approximately proportional, and stronger
fractional Mg~II responses relative to the continuum. Values consistent with zero indicate no measurable coupling, while negative
values indicate anti-correlated behavior. DN~Tau shows the clearest evidence for an anti-correlation, while
SU~Aur shows a weaker indication; the other negative slopes are consistent with zero.

The observations therefore reveal substantial object-to-object diversity rather than a universal temporal relation between Mg~II and
the near-UV continuum. This is consistent with the composite origin of Mg~II emission: chromospheric, magnetospheric, and outflow
components may vary differently, while optical-depth changes or wind absorption can modify the observed line flux independently of the
continuum. A qualitatively similar diversity is seen in the ULLYSES monitoring of four CTTSs \citep{2024ApJ...970..118W}, whose
Mg~II--accretion luminosity relations differ markedly among sources. Thus, both data sets indicate that population-level relations do not
necessarily translate into a uniform  temporal response within individual TTSs.

\begin{figure}
\centering
\includegraphics[width=11cm]{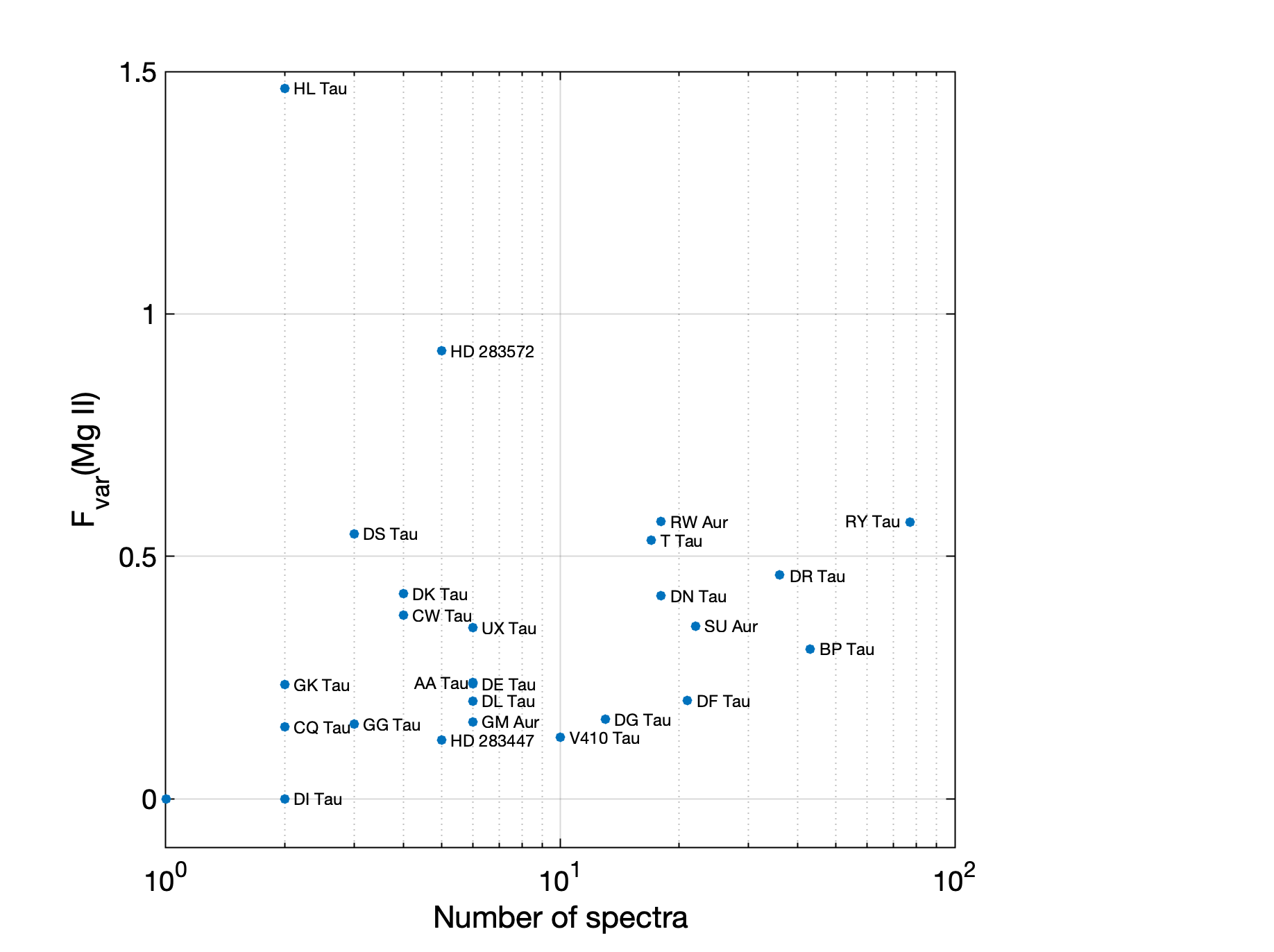}
\caption{Fractional variability amplitude, $F_{\rm var}$, of the Mg~II emission as a function of the number of available IUE spectra.
$F_{\rm var}$ represents the variability amplitude after accounting for the measurement uncertainties.}
\end{figure}

\begin{figure}
\centering
\includegraphics[width=10cm]{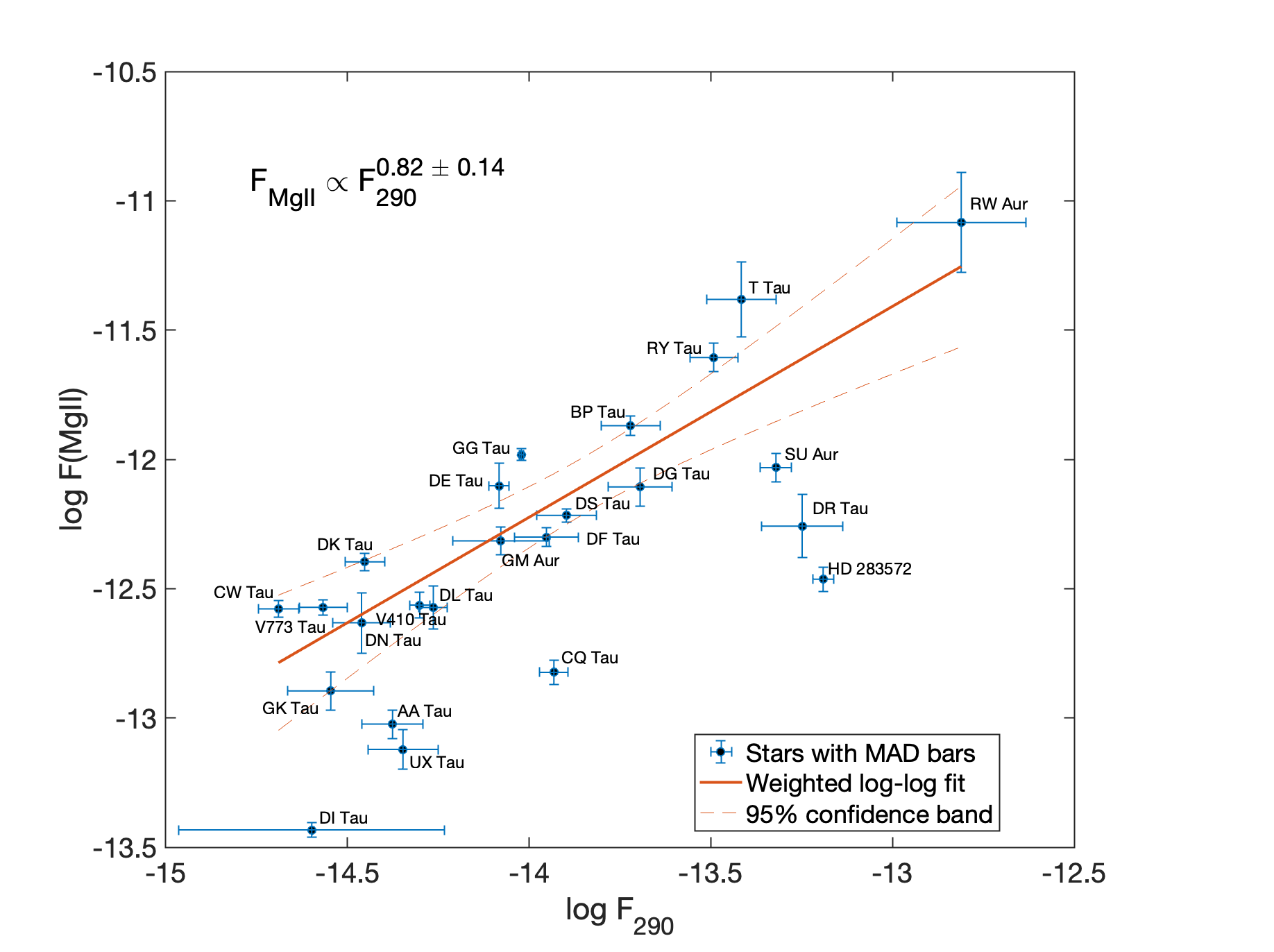}
\caption{Mean continuum-subtracted Mg~II flux versus mean UV continuum flux at 2900~\AA. The solid line shows the weighted best-fitting
relation, $F({\rm MgII})\propto F_{2900}^{0.82\pm0.14}$, and the dashed curves its 95\% confidence interval. Error bars are based on
the median absolute deviation (MAD).}
\end{figure}

\begin{figure}
\includegraphics[width=9cm]{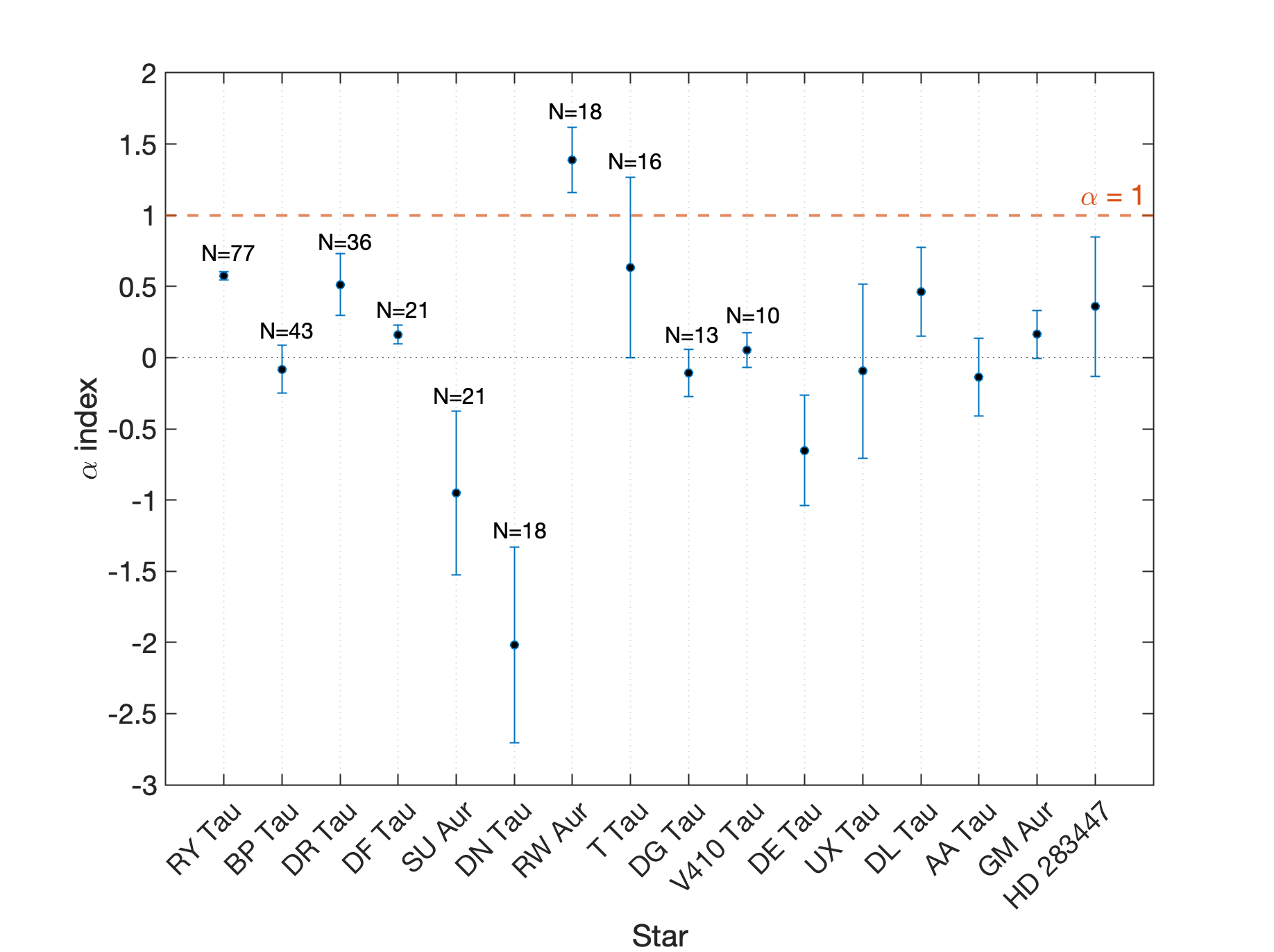}
\caption{Power-law index, $\alpha$, of the Mg~II--continuum relation for individual stars. Error bars show the formal regression
uncertainties, and labels give the number of spectra. The dashed line marks the proportional relation ($\alpha=1$).}
\end{figure}

\begin{table*}
\centering
\caption{Power-law fits between the net Mg II flux and the UV continuum for individual sources.}
\label{tab:mgii_individual_fits}
\begin{tabular}{lrrrrrr}
\hline
Object & $N_{\rm spec}$ & $\alpha$ & $\beta$ & $r_{\rm P}$ & $r_{\rm S}$ & rms \\
\hline
RY Tau    & 77 & $0.57 \pm 0.03$  & $-3.84 \pm 0.41$  &  0.91 &  0.79 & 0.05 \\
BP Tau    & 43 & $-0.08 \pm 0.17$ & $-13.02 \pm 2.32$ & -0.08 &  0.04 & 0.15 \\
DR Tau    & 36 & $0.51 \pm 0.22$  & $-5.49 \pm 2.88$  &  0.38 &  0.43 & 0.21 \\
DF Tau    & 21 & $0.16 \pm 0.07$  & $-10.04 \pm 0.93$ &  0.49 &  0.40 & 0.07 \\
SU Aur    & 21 & $-0.95 \pm 0.57$ & $-24.74 \pm 7.64$ & -0.36 &  0.03 & 0.18 \\
DN Tau    & 18 & $-2.02 \pm 0.69$ & $-42.00 \pm 9.95$ & -0.59 & -0.07 & 0.43 \\
RW Aur    & 18 & $1.39 \pm 0.23$  & $6.64 \pm 2.95$   &  0.83 &  0.84 & 0.26 \\
T Tau     & 16 & $0.63 \pm 0.63$  & $-2.97 \pm 8.52$  &  0.26 &  0.41 & 0.29 \\
DG Tau    & 13 & $-0.11 \pm 0.17$ & $-13.62 \pm 2.28$ & -0.20 & -0.16 & 0.08 \\
V410 Tau  & 10 & $0.05 \pm 0.12$  & $-11.81 \pm 1.76$ &  0.15 &  0.09 & 0.07 \\
DE Tau    &  6 & $-0.65 \pm 0.39$ & $-21.31 \pm 5.46$ & -0.64 & -0.31 & 0.08 \\
UX Tau    &  6 & $-0.10 \pm 0.61$ & $-14.52 \pm 8.78$ & -0.08 & -0.31 & 0.17 \\
DL Tau    &  6 & $0.46 \pm 0.31$  & $-6.01 \pm 4.45$  &  0.59 &  0.54 & 0.07 \\
AA Tau    &  6 & $-0.14 \pm 0.27$ & $-15.00 \pm 3.92$ & -0.24 & -0.43 & 0.13 \\
GM Aur    &  6 & $0.16 \pm 0.17$  & $-10.02 \pm 2.36$ &  0.44 &  0.60 & 0.06 \\
HD 283447 &  5 & $0.36 \pm 0.49$  & $-7.36 \pm 7.14$  &  0.39 &  0.40 & 0.06 \\
\hline
\end{tabular}
\end{table*}

\vspace{0.5cm}
\noindent
{\it Possible contribution from the RY~Tau outflow}
\vspace{0.2cm}

The IUE spectra were obtained through the $10''\times20''$ large aperture, with a spatial resolution of only $\sim2''$--$3''$.
Emission from the inner RY~Tau outflow may therefore contribute to the unresolved Mg~II flux. RY~Tau, one of the best-sampled objects in
our data set, provides an opportunity to examine this possibility.

RY~Tau drives an optical jet \citep{1994ApJS...93..485H}, and hot plasma associated with its inner outflow has been detected in the
C~III] and Si~III] UV lines \citep{2007ApJ...654L..91G}. High-resolution GHRS and STIS spectra obtained several years apart
showed little variability in the blueshifted emission attributed to the wind, which was interpreted as arising from a stellar-driven
outflow associated with open magnetic-field structures.

An outflow contribution to Mg~II can be represented phenomenologically as
\begin{equation}
F_{\rm MgII}(t) =
A F_{\rm cont}(t)^{\alpha} + F_{\rm out}(t),
\end{equation}
where $F_{\rm out}(t)$ represents emission not correlated with the near-UV continuum. If this component varies more slowly than the
continuum-correlated emission, as suggested by the C~III] and Si~III] observations, it could produce a temporally coherent
signature in the residuals.

We therefore examined the residuals of the RY~Tau Mg~II--continuum relation after excluding the isolated highest-continuum observation,
which strongly affects the fit. No persistent offset or systematic temporal trend is apparent, and the residuals change sign even within
the densely sampled intervals. Thus, although an unresolved outflow contribution to Mg~II cannot be excluded, the IUE data provide no
evidence for a temporally distinct component that can be separated from the continuum-correlated emission.

\section{Time-domain analysis}

The temporal sampling of the IUE observations is highly
non-uniform (see Appendix~B). Figure~8 shows intervals between
consecutive observations concentrated from fractions of a day to
several days, corresponding to observing campaigns, together with
much longer gaps. A relative minimum at $\Delta t\simeq18.2$~d was
adopted as an empirical threshold for separating observations within
and between campaigns. The archive therefore combines relatively
dense short-term sampling with long, irregularly sampled baselines.

\begin{figure}
\centering
\includegraphics[width=7.5cm]{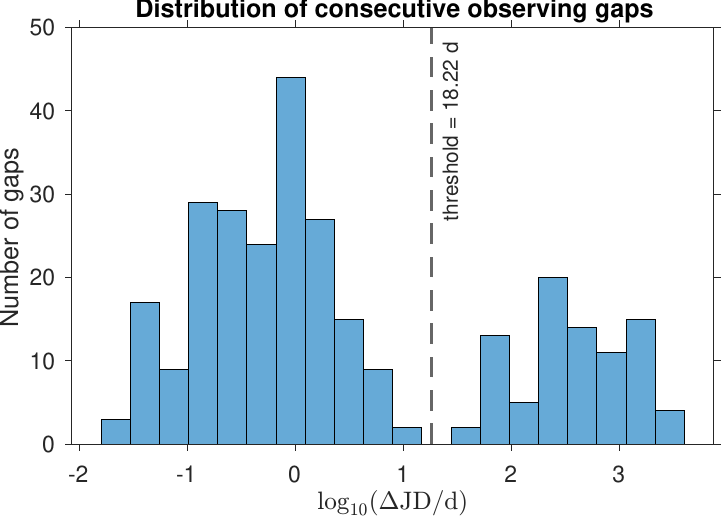}
\caption{Distribution of time intervals, $\Delta{\rm JD}$, between
consecutive IUE observations. The dashed line at
$\Delta{\rm JD}=18.22$~d marks the empirical threshold adopted to
separate observing campaigns.}
\end{figure}

Before searching for recurrent signals, we examined the light curves
for isolated high-amplitude events that could affect the variance and
periodograms. Candidate continuum events were identified independently
in $F_{2700}$, $F_{2900}$, $F_{3030}$, and the continuum beneath
Mg~II, and retained only when at least three of the four diagnostics
deviated simultaneously in the same direction. Each candidate was
then checked against the original IUE spectrum to exclude instrumental
effects.

Because the observations are clustered, we also tested whether the
median and MAD should be estimated after collapsing individual
observing campaigns. With the 18.2-d campaign
definition, however, this leaves only six campaign values for DF~Tau
and eight for BP~Tau, producing poorly constrained MAD thresholds.
Moreover, campaign medians can suppress the short-lived excursions
that this analysis is intended to identify. We therefore retain the
observation-based median/MAD criterion together with the
multi-diagnostic requirement above.

This procedure identified a saturated RY~Tau exposure, which was
discarded, leaving only two reliable continuum outliers, in DF~Tau
and BP~Tau (Fig.~9. Both represent relatively modest enhancements.
The IUE observations therefore provide little evidence for
large-amplitude continuum bursts, although rare events could be
missed by the irregular sampling.

Mg~II shows larger excursions and more statistical outliers than the
continuum, most of them in BP~Tau and generally without simultaneous
continuum brightenings. This behavior points to variability in the
Mg~II-emitting plasma that is at least partly decoupled from the
continuum-producing regions.

BP~Tau has a strong, predominantly axisymmetric large-scale
magnetic field, with accretion concentrated at high latitudes
\citep{2008MNRAS.386.1234D}, and rotational modulation has been
reported in accretion-related diagnostics
\citep{1990AJ....100.1957S}. The Mg~II excursions found here,
however, generally lack comparable continuum variations, suggesting
that the line-emitting plasma can vary partly independently even in
a system with a relatively ordered large-scale magnetosphere. The
IUE data alone cannot establish whether this behavior is related to
the magnetic topology.

\begin{figure*}
\centering
\includegraphics[width=15cm]{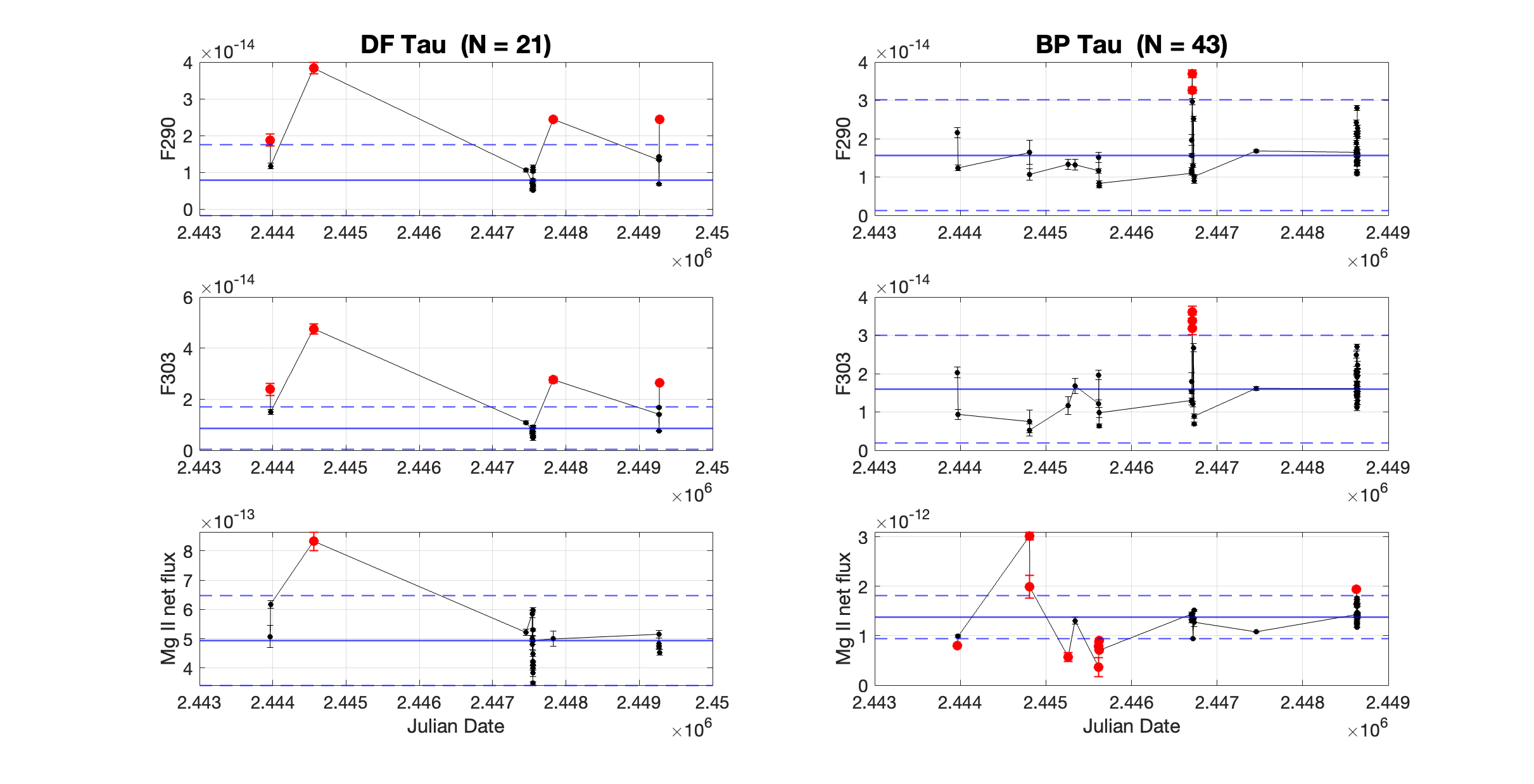}
\caption{Continuum and Mg~II variability and outliers in DF~Tau
(left) and BP~Tau (right). From top to bottom, the panels show the
continuum fluxes at 2900 and 3030~\AA\ and the continuum-subtracted
Mg~II flux as a function of Julian Date. Solid blue lines indicate
the median flux and dashed blue lines the adopted MAD-based outlier
thresholds. Red symbols mark measurements classified as outliers.}
\end{figure*}

These results reinforce the diversity of the Mg~II--continuum
behavior discussed above and provide context for the period search
below.

\subsection{Time-series analysis}

The IUE observations are highly irregular and often clustered into short campaigns separated by long gaps (Fig.~8). We therefore
used the Lomb--Scargle periodogram to search for periodic or recurrent signals in the unevenly sampled time series.

The analysis was restricted to objects with at least 20 observations, a pragmatic threshold that excludes extremely sparse time series but
does not guarantee adequate temporal coverage. Three observables were analyzed: ({\it i}) the UV continuum interpolated beneath Mg~II,
({\it ii}) the continuum-subtracted Mg~II flux, and ({\it iii}) the Mg~II-to-continuum flux ratio.

Times were expressed relative to the first observation of each star and the mean was subtracted from each series. No detrending was
applied because the clustered sampling does not constrain secular variations reliably. Periodograms were evaluated over periods from
0.5~d to the smaller of 300~d or one third of the total baseline. This upper limit defines only the frequency range explored and does
not imply that periods approaching it are adequately sampled.

Rotationally modulated accretion signatures may also lose phase coherence between observing campaigns as accretion structures evolve.
Individual campaigns generally contain too few observations or cover too short an interval for independent period determinations.
Consequently, we analyse the complete time series while recognizing that significant periodogram features need not establish periodicity
or a characteristic physical timescale.

The significance of the highest peak was estimated by 1000 non-parametric bootstrap realizations in which the fluxes were
randomly permuted while keeping the observing times fixed. The false-alarm probability (FAP) was calculated as the fraction of
realizations whose maximum Lomb--Scargle power exceeded the observed maximum. Significant features were further assessed against the
spectral window and by inspection of the phase-folded light curves and their effective temporal coverage.

The periodograms show prominent peaks near 33~d in the continua of DR~Tau and DF~Tau and near 10~d in the Mg~II flux of DR~Tau. These
features satisfy the bootstrap significance criterion and are not obviously reproduced in the corresponding spectral windows
(Fig.~10). Their interpretation is nevertheless limited by the temporal sampling, and Fig.~10 distinguishes observations from the
densest campaign from those obtained at widely separated epochs to show the effective phase coverage.

BP~Tau also shows a formally significant continuum peak near 15~d (${\rm FAP}=0.003$), approximately twice the commonly adopted
rotation period of 7.6~d \citep{1990AJ....100.1957S}. Earlier IUE monitoring showed that rotational modulation in BP~Tau can lose
coherence and differs among UV diagnostics \citep{1997ApJ...482..465G}; we therefore do not interpret the 15-d feature as a rotation period.

\subsubsection{DR Tau}

DR~Tau is a strongly accreting and highly variable CTTS for which no unique rotation period is established. Published estimates range
from a few days to $\sim10$~d, while recent high-cadence photometry favors a period near 3~d and shows that rapidly evolving accretion
structures and the nearly pole-on geometry complicate the recovery of stable rotational signals \citep{2025A&A...699A.221Z}.

The periodograms show formally significant peaks at 10.01~d in Mg~II and 33.46~d in the UV continuum, neither of which is obviously
reproduced by the corresponding spectral window (Fig.~10). However, the densest observing campaign spans only $\sim13$~d, corresponding
to about 1.3 cycles of the Mg~II feature and 0.4 cycles of the continuum feature. The remaining phase coverage comes from widely
separated campaigns, over which the accretion geometry and variability pattern may have evolved. We therefore report the 10- and 33-d peaks
as formally significant periodogram features, rather than as detections of periodic or characteristic timescales, with the 33-d
feature particularly poorly constrained.

\subsection{DF Tau}

The DF~Tau continuum periodogram shows a formally significant
peak at 33.01~d that is not obviously reproduced by the spectral
window (Fig.~10). DF~Tau is a close binary with an orbital period of
$\sim44$~yr \citep{2014AJ....147..157S}, ruling out binary orbital
motion as the origin of the 33-d feature; however, the unresolved IUE
data do not identify which component dominates the variability.
Although the phase-folded light curve appears coherent, the densest
observing sequence spans only $\sim14$~d, or about 0.4 cycles of the
trial period, with the remaining phase coverage provided by widely
separated campaigns. We therefore report the 33.01-d peak as a
formally significant periodogram feature, but do not interpret it as
evidence for a periodic or characteristic physical timescale.

\begin{figure*}
\includegraphics[width=18cm]{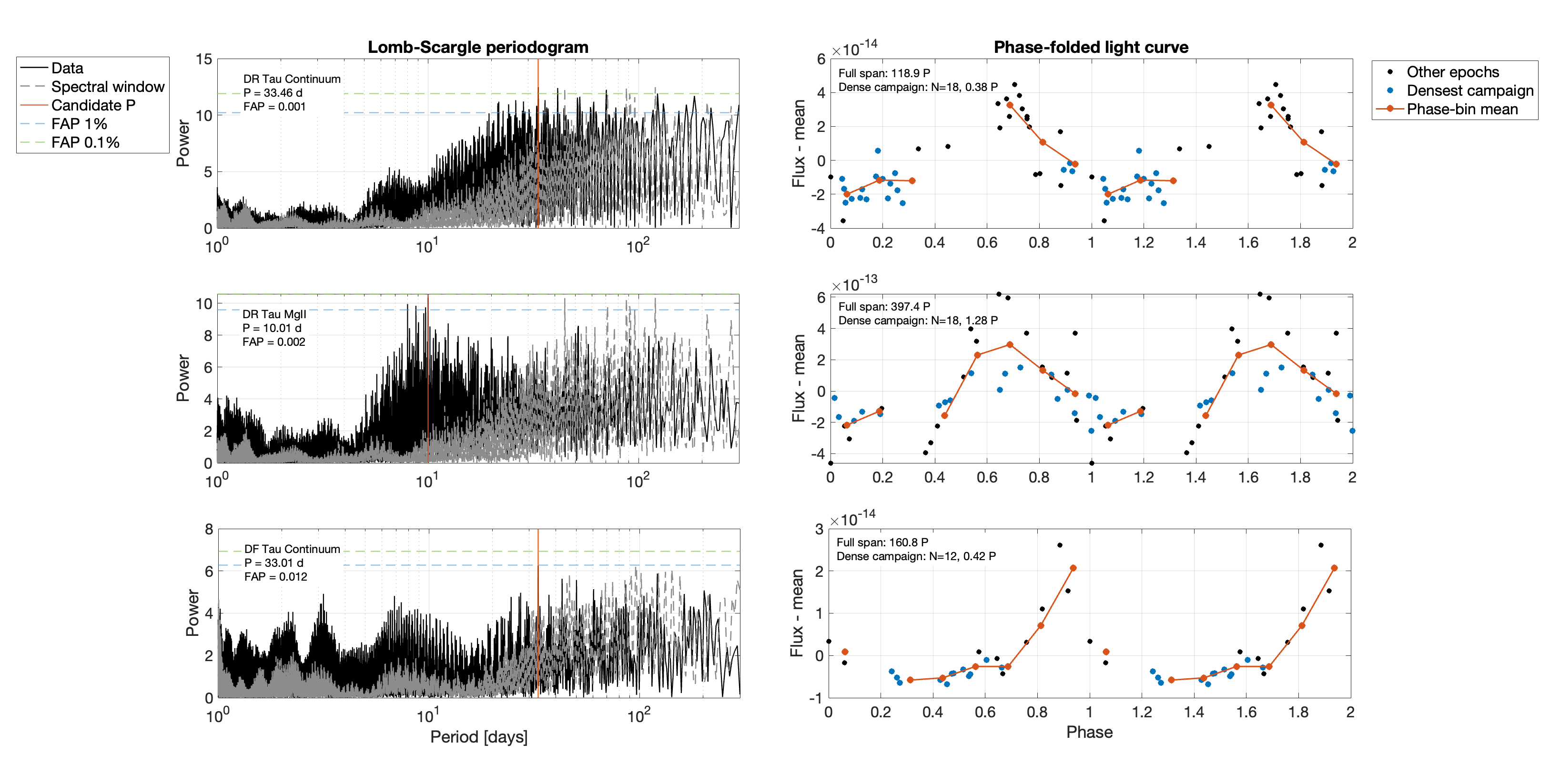}
\caption{Lomb--Scargle periodograms and phase-folded light curves for DR~Tau and DF~Tau.
Left: periodograms for the DR~Tau continuum, DR~Tau Mg~II flux, and DF~Tau continuum. gray curves show the corresponding spectral windows,
red vertical lines mark the selected peaks, and horizontal dashed lines the 1\% and 0.1\% bootstrap false-alarm thresholds.
Right: light curves folded at the corresponding periods; colored symbols identify observations from the densest campaign and orange
curves show phase-binned averages. The dense campaigns span $\sim0.4$ cycles of the 33-d features and $\sim1.3$ cycles of the
10.01-d feature.}

\end{figure*}

Taken together, these results illustrate the limitations of period searches in the irregularly sampled IUE archive. Formally significant Lomb--Scargle peaks do not necessarily establish recurrent signals when the effective phase coverage is limited. We therefore report the identified peaks as periodogram features rather than detections of periodic or characteristic timescales.

The occurrence of two continuum features near 33~d also prompted us to consider a possible spacecraft-related origin. The approximately one-sidereal-day orbital period of IUE cannot directly account for these features. However, \citet{1993PASP..105..603A} identified persistent 28--33-d modulations in the charged-particle flux measured by IUE during perigee passages and associated them with recurrent high-speed solar-wind streams. This result does not establish that the stellar UV flux measurements are modulated at the same periods, but an environmental or instrumental contribution cannot be excluded. Together with the limited cycle coverage, this possibility provides an additional reason not to assign an astrophysical interpretation to the two $\sim33$-d features.

\section {Summary and conclusions}

We have investigated the long-term near-UV variability of TTSs in the Taurus--Auriga star-forming region using the IUE archive. The
principal results are:

\begin{itemize}

\item The near-UV continuum is strongly variable, while the continuum flux ratios between 2675 and 3055~\AA\ show comparatively small
variations in most objects. Thus, within this wavelength range, variability is dominated by changes in continuum level rather than
large changes in spectral shape.

\item The near-UV and quasi-simultaneous optical fluxes are significantly correlated in six of the seven best-sampled sources,
although the strength and relative amplitudes of the variations differ among objects. The UV--optical behavior therefore cannot be described
by a single proportional relation, suggesting different relative contributions from accretion-related emission, stellar variability,
extinction, and geometry.

\item Mg~II shows substantial object-to-object diversity in its relation to the near-UV continuum. Some sources show correlated
variations with different fractional responses, whereas in others the two diagnostics are largely decoupled. DN~Tau shows evidence for an
anti-correlation, with weaker evidence in SU~Aur. There is therefore no universal temporal response of Mg~II to continuum variability,
consistent with contributions from chromospheric, magnetospheric, and outflowing plasma.

\item The IUE time series are highly irregular and clustered into observing campaigns separated by long gaps. Formally significant
periodogram features are found in some of the best-sampled objects, but their interpretation is limited by the spectral window and
effective phase coverage. We therefore do not claim detections of periodic or characteristic timescales.

\end{itemize}

The IUE archive thus reveals substantial source-to-source diversity in the long-term near-UV variability of TTSs. Population-level
relations between UV diagnostics do not necessarily translate into a common temporal behavior within individual objects. Despite the
higher sensitivity and spectral resolution of modern UV facilities, the combination of repeated UV spectroscopy, long temporal baselines,
and quasi-simultaneous optical measurements makes the IUE archive uniquely valuable for long-term variability studies.

\begin{acknowledgements}

This research was partially supported by grant PID2023-147740OB-I00 from the Ministry of Science, Innovation and Universities  of Spain. 
 \end{acknowledgements}

\bibliographystyle{aa} 
\bibliography{references} 

\newpage
\begin{appendix}

\section{Near-UV continuum relations for the individual sources}
\label{app:continuum}

The continuum flux measured in the 2900 \AA\ window is plotted against the fluxes in the 2700 \AA\ and 3030 \AA\ windows  for  the TTSs with sufficient numbers of observations (N > 10), including BP~Tau, RW~Aur, DR~Tau, T~Tau, DF~Tau, and SU~Aur. 

\begin{figure*}[p]
\centering

\includegraphics[width=0.29\textwidth]{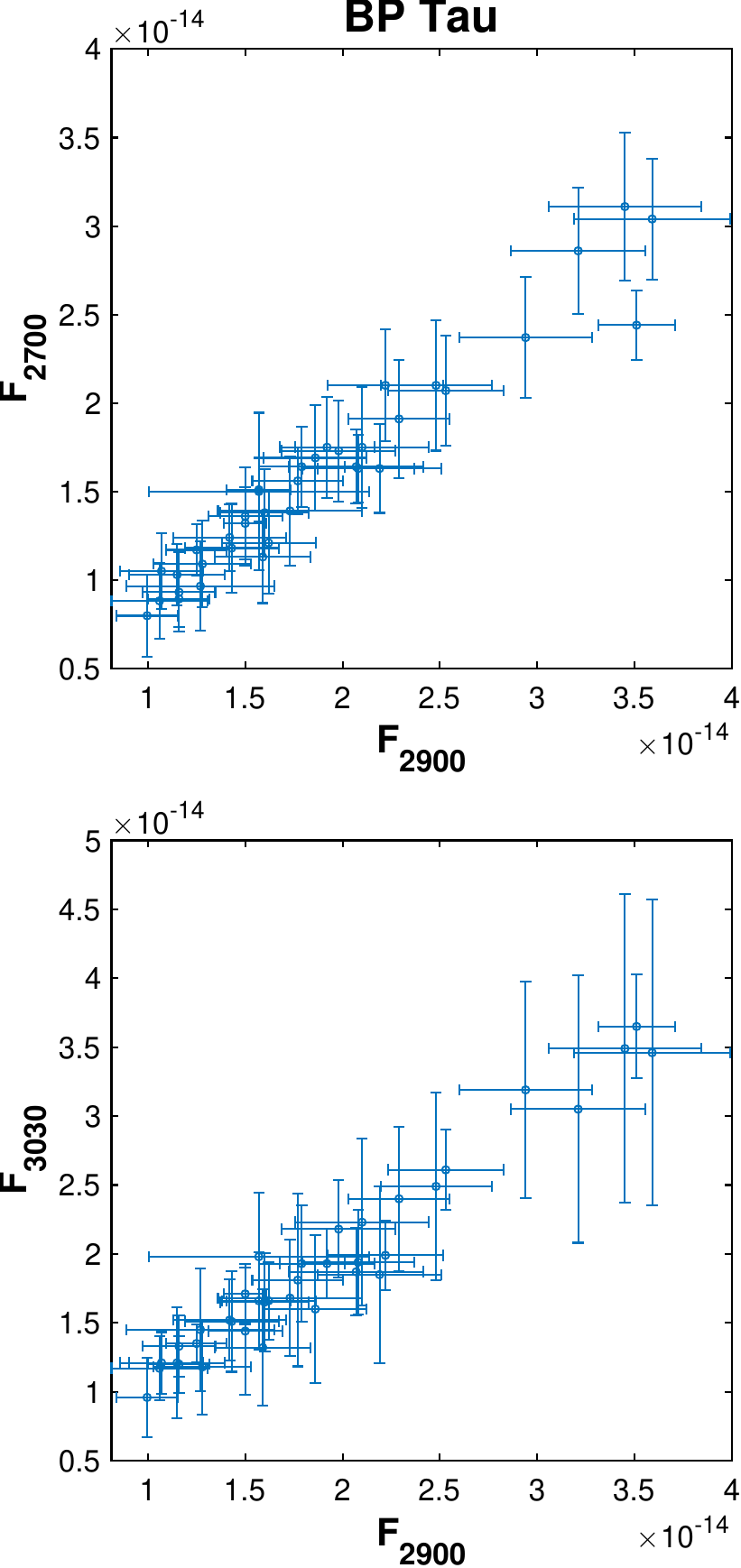}\hfill
\includegraphics[width=0.2805\textwidth]{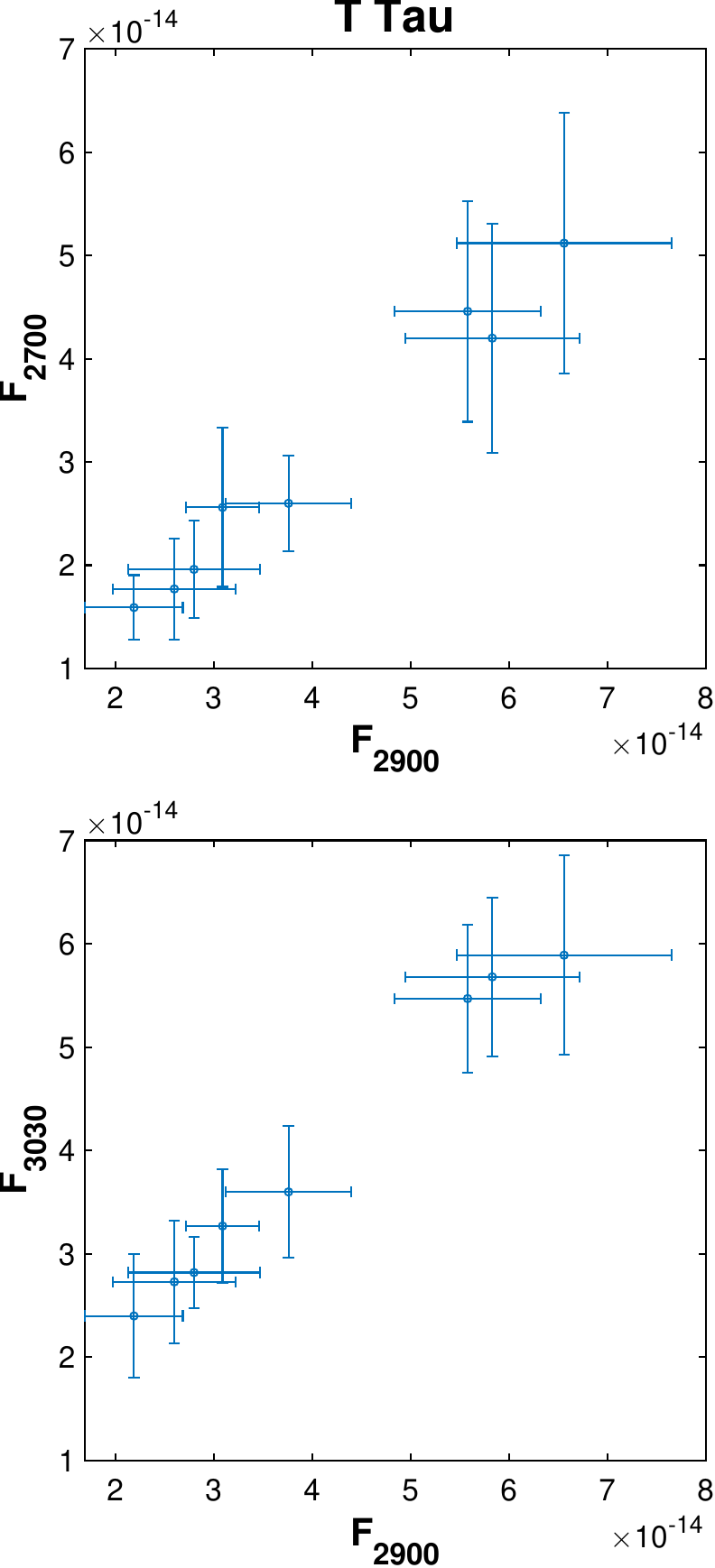}\hfill
\includegraphics[width=0.29\textwidth]{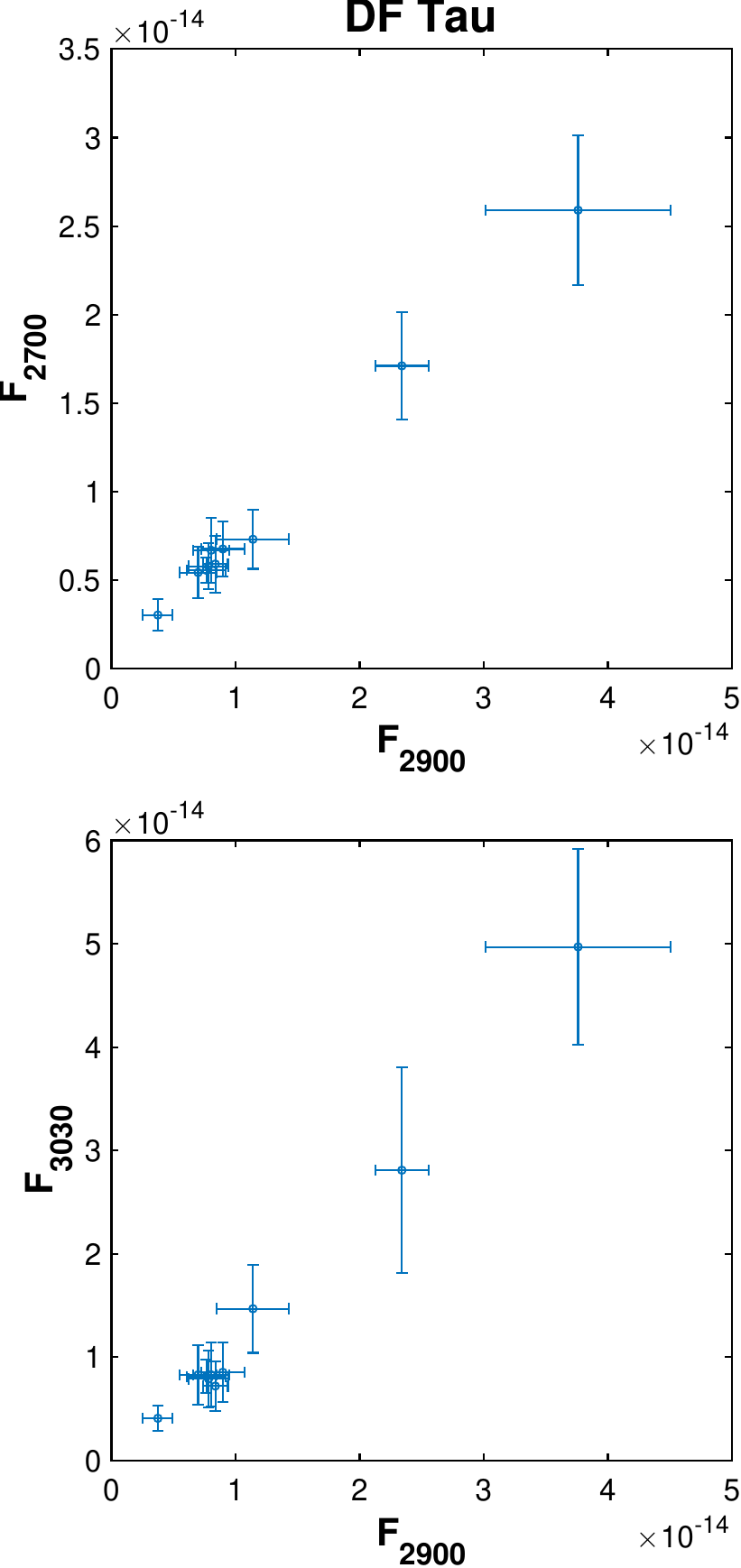}

\vspace{0.3cm}

\includegraphics[width=0.29\textwidth]{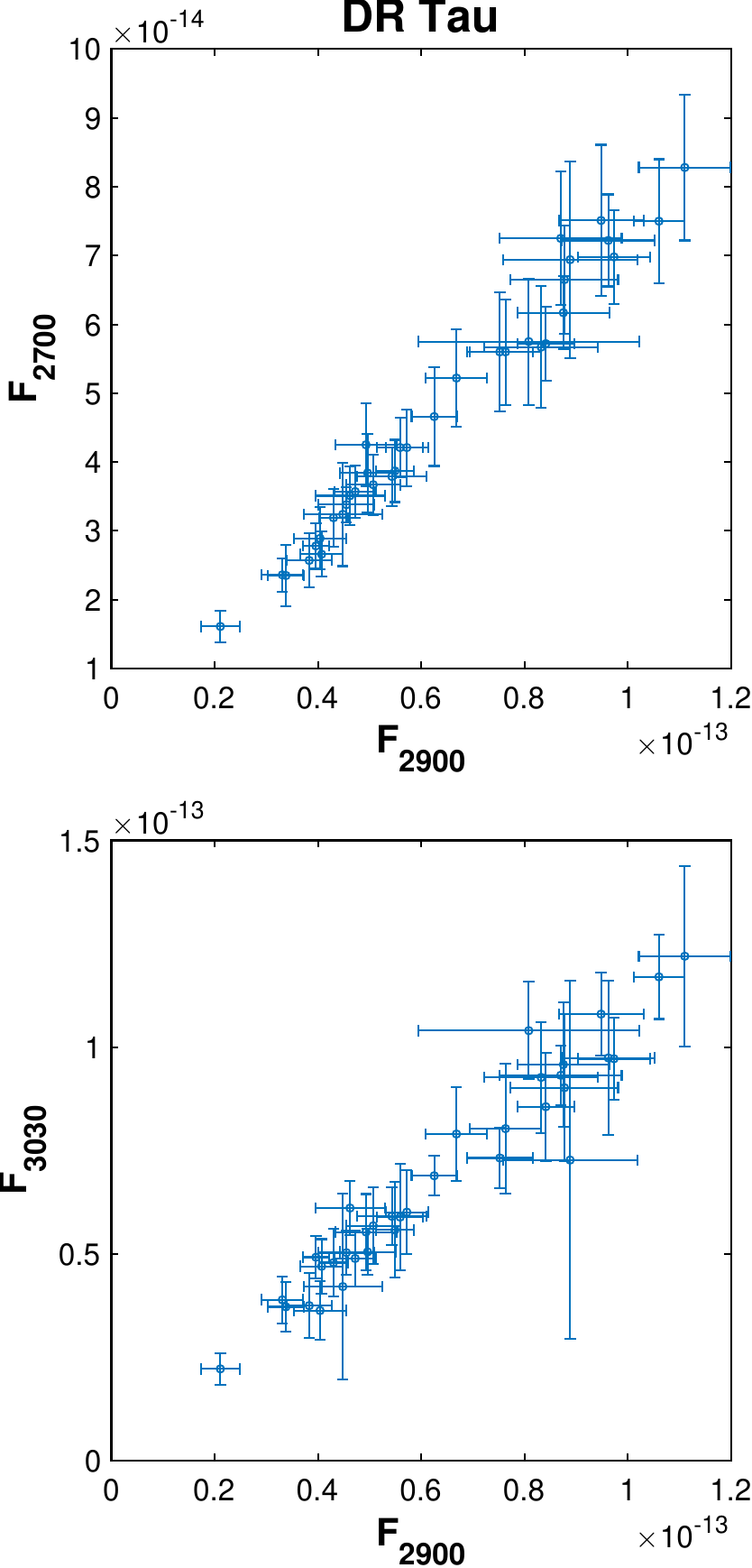}\hfill
\includegraphics[width=0.29\textwidth]{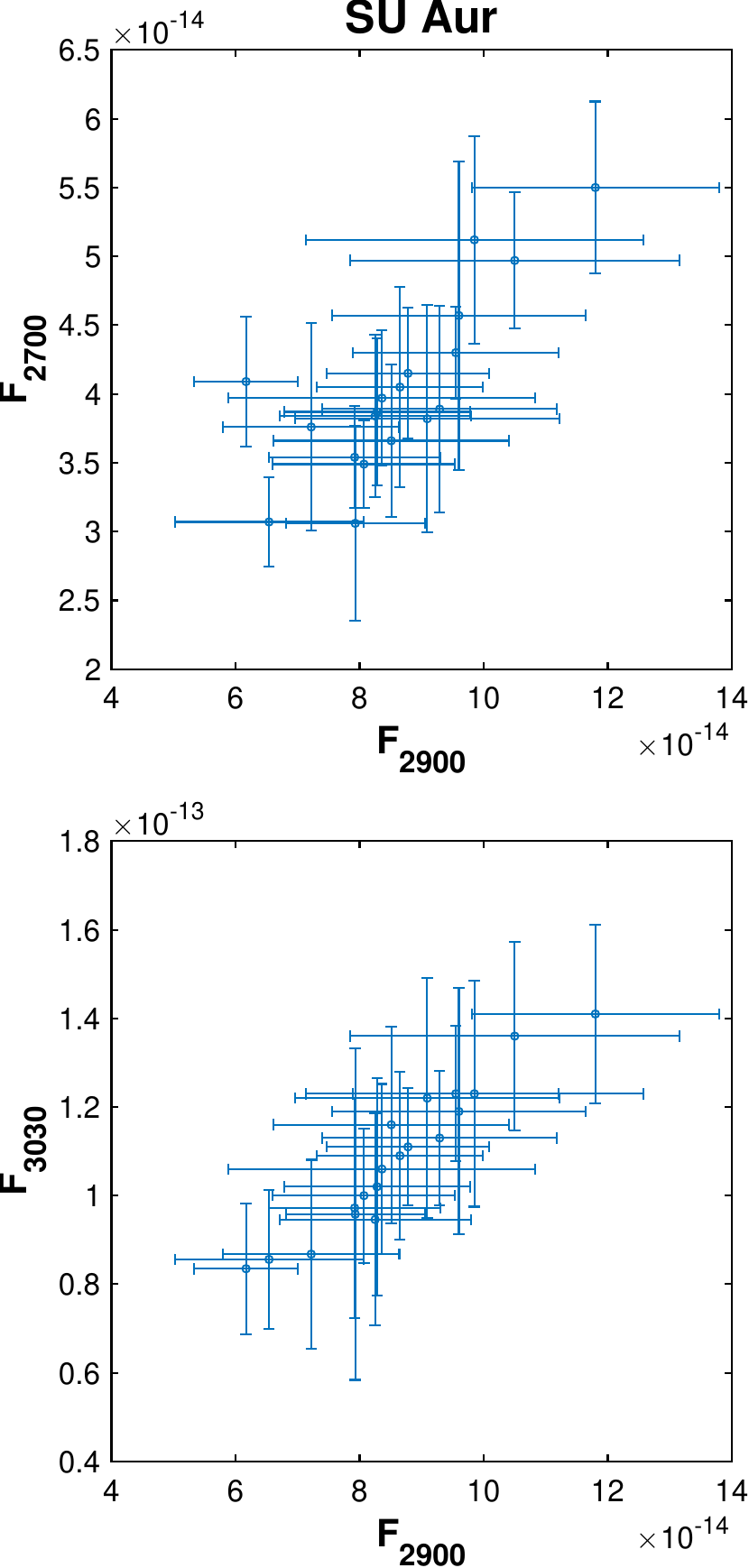}\hfill
\includegraphics[width=0.29\textwidth]{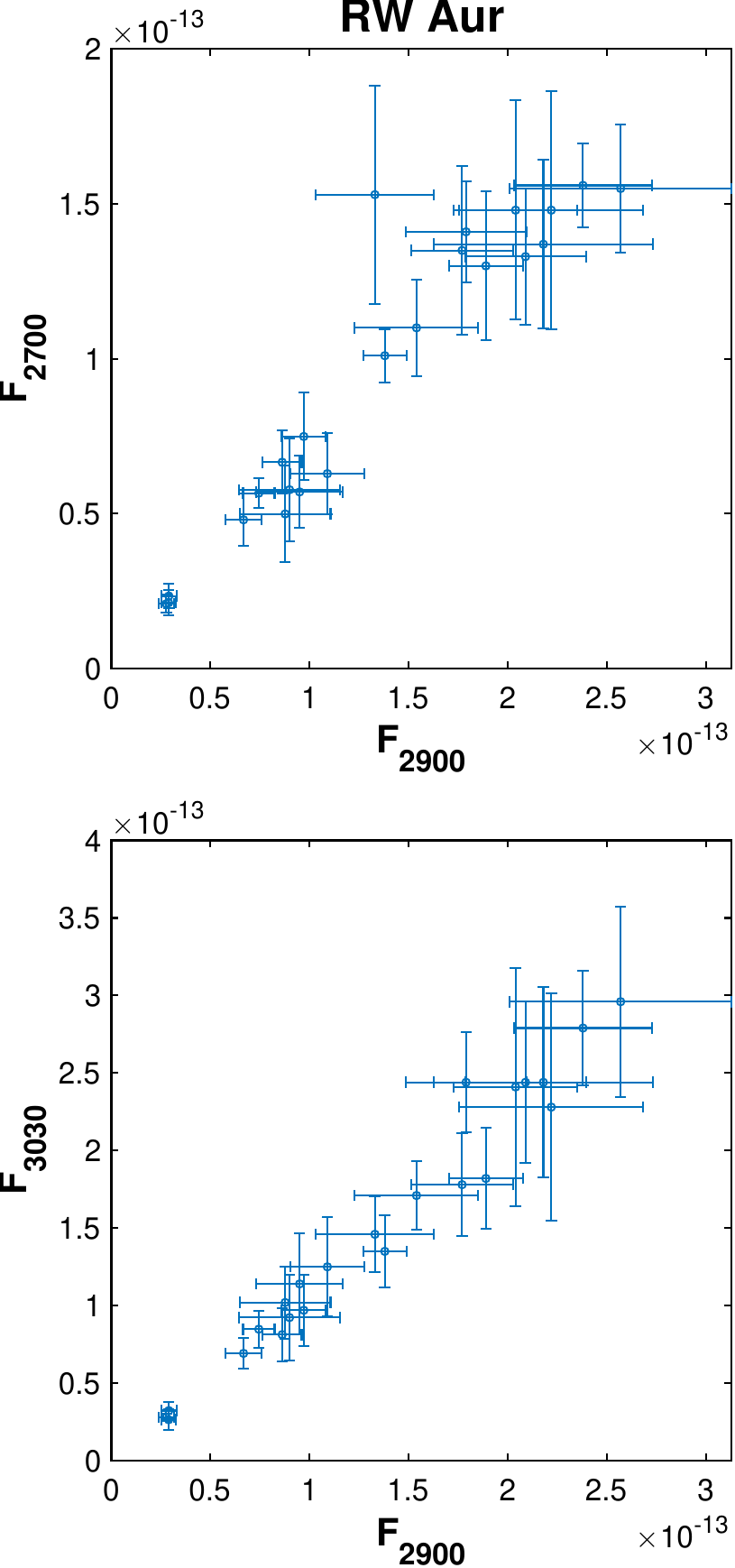}

\caption{Relations between the near-UV continuum fluxes measured in
the three continuum windows for the individual sources. For each
source, the upper panel shows $F_{2700}$ as a function of $F_{2900}$,
while the lower panel shows $F_{3030}$ as a function of $F_{2900}$.
Error bars represent the uncertainties associated with the individual
flux measurements.}
\label{fig:app_continuum}
\end{figure*}

\section{Temporal sampling and near-UV and Mg~II variability of the individual sources} 
\label{app:timesampling}

\begin{figure*} 
\centering
 \includegraphics[width=0.45\textwidth]{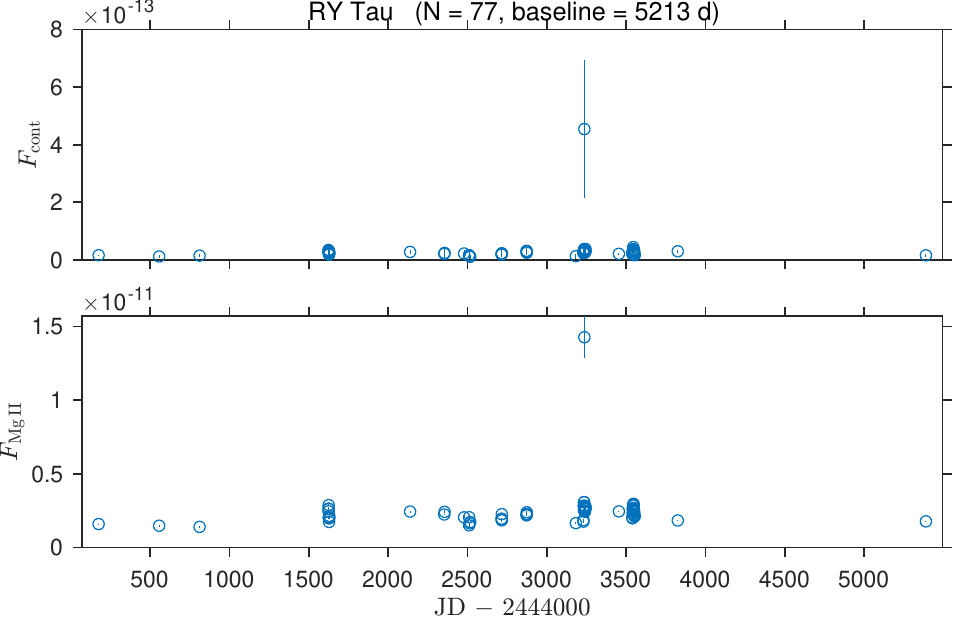} 
\vspace{0.25cm} 
 \includegraphics[width=0.45\textwidth]{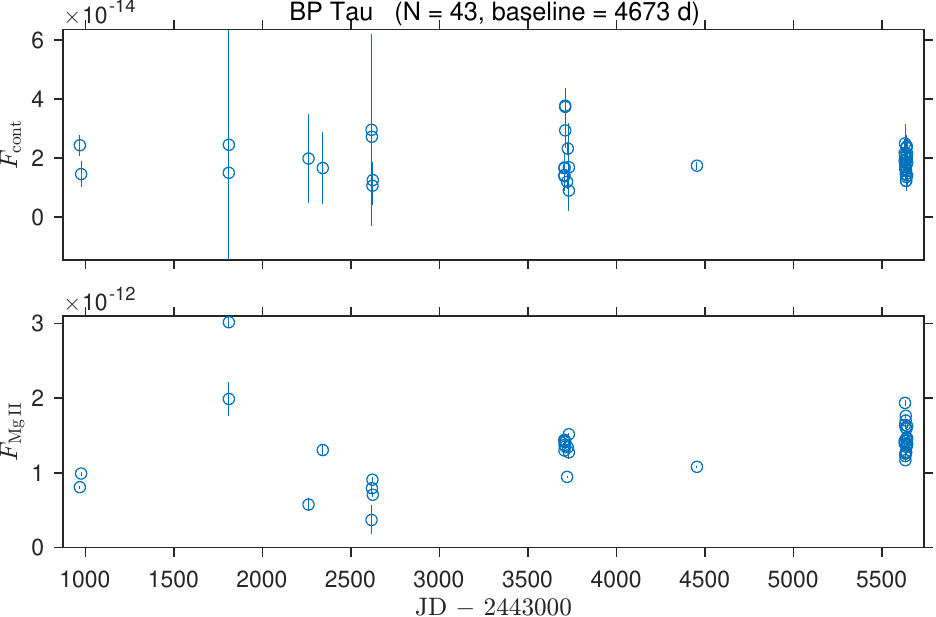} 
\vspace{0.25cm} 
 \includegraphics[width=0.45\textwidth]{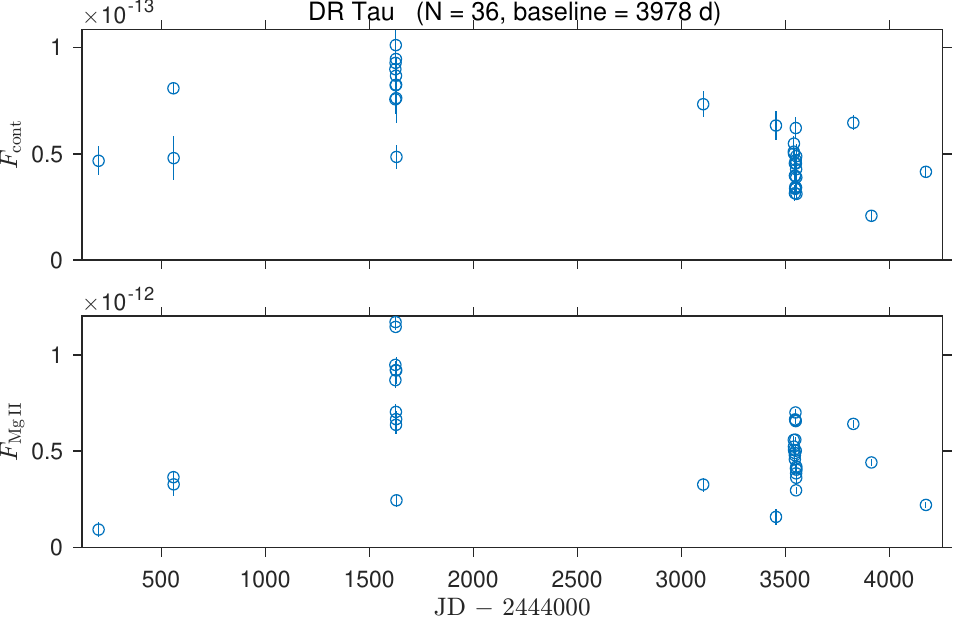} 
\vspace{0.25cm}
  \includegraphics[width=0.45\textwidth]{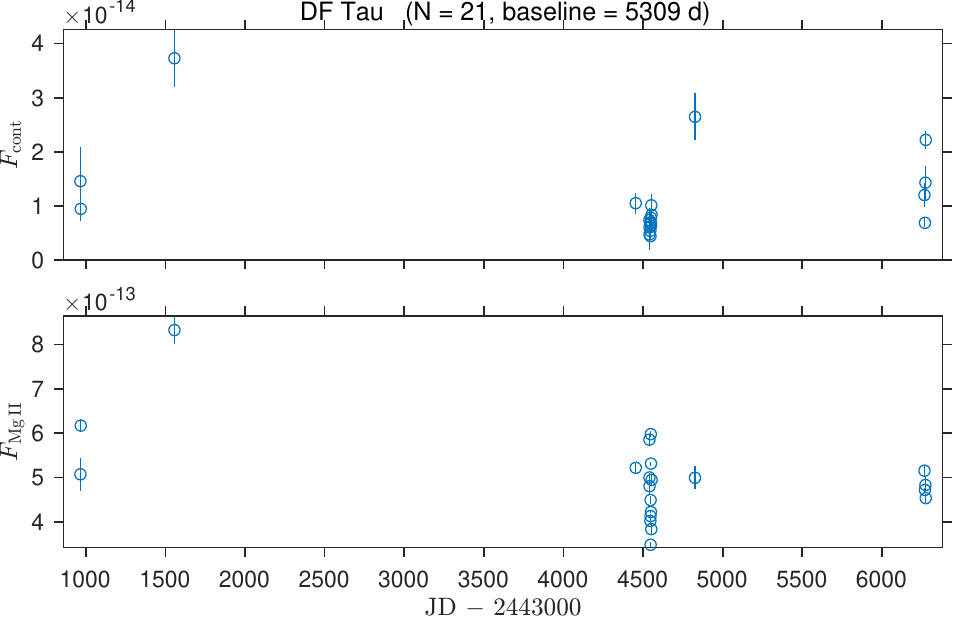} 
\vspace{0.25cm} 
  \includegraphics[width=0.45\textwidth]{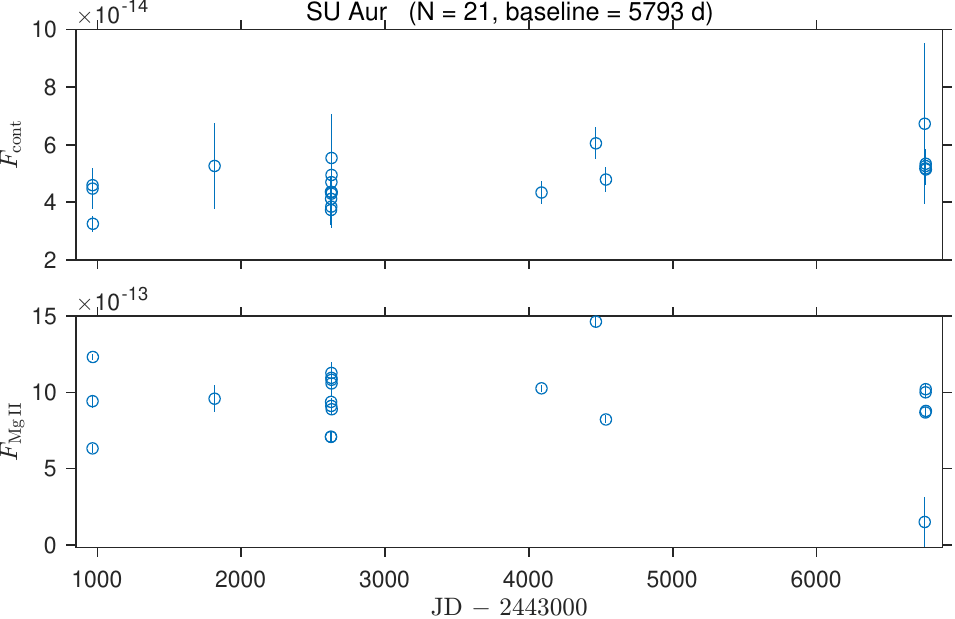} 
\vspace{0.25cm} 
  \includegraphics[width=0.45\textwidth]{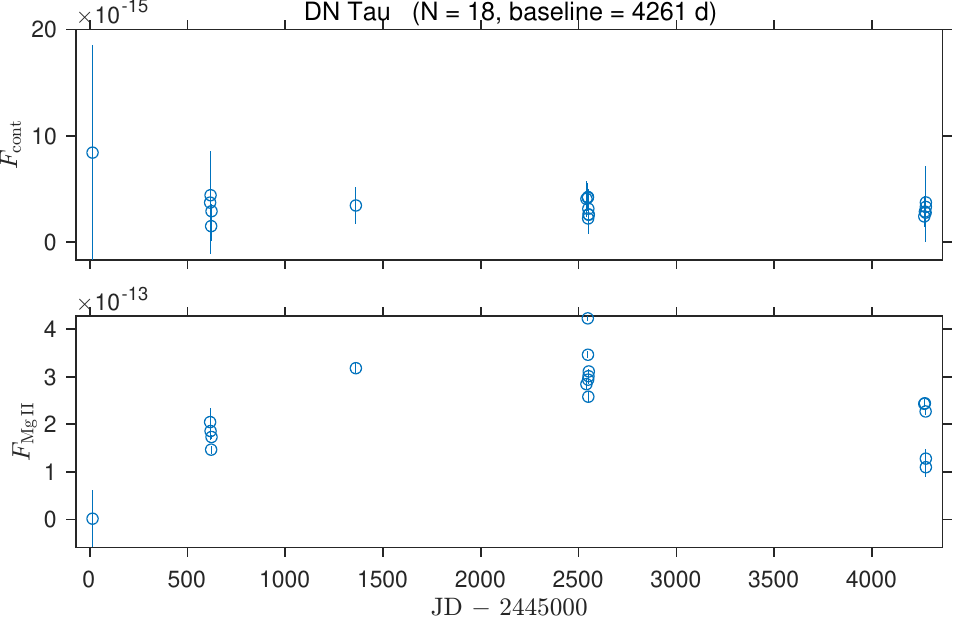} 
\vspace{0.25cm} 
  \includegraphics[width=0.45\textwidth]{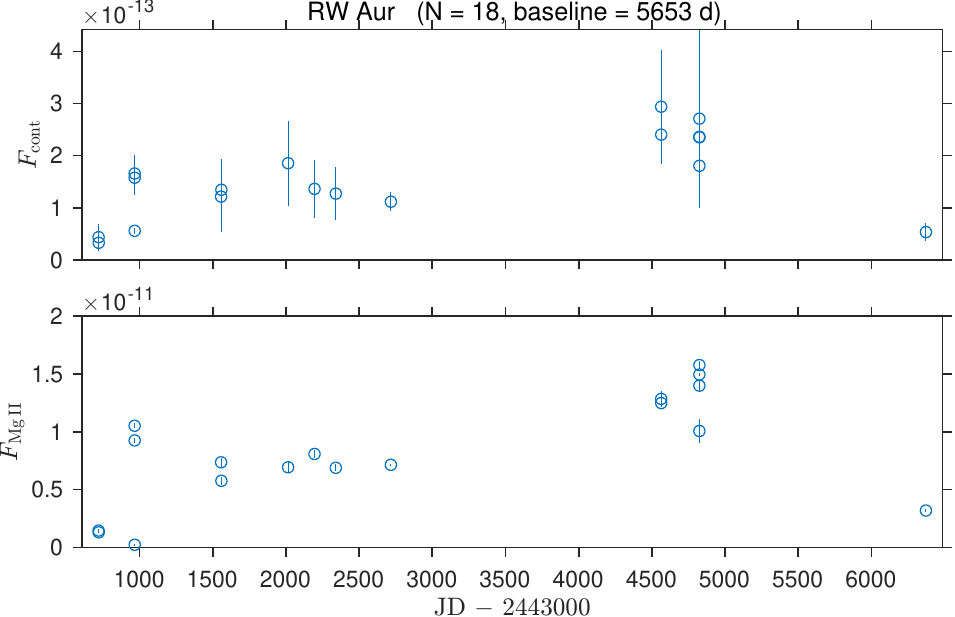} 
\vspace{0.25cm} 
  \includegraphics[width=0.45\textwidth]{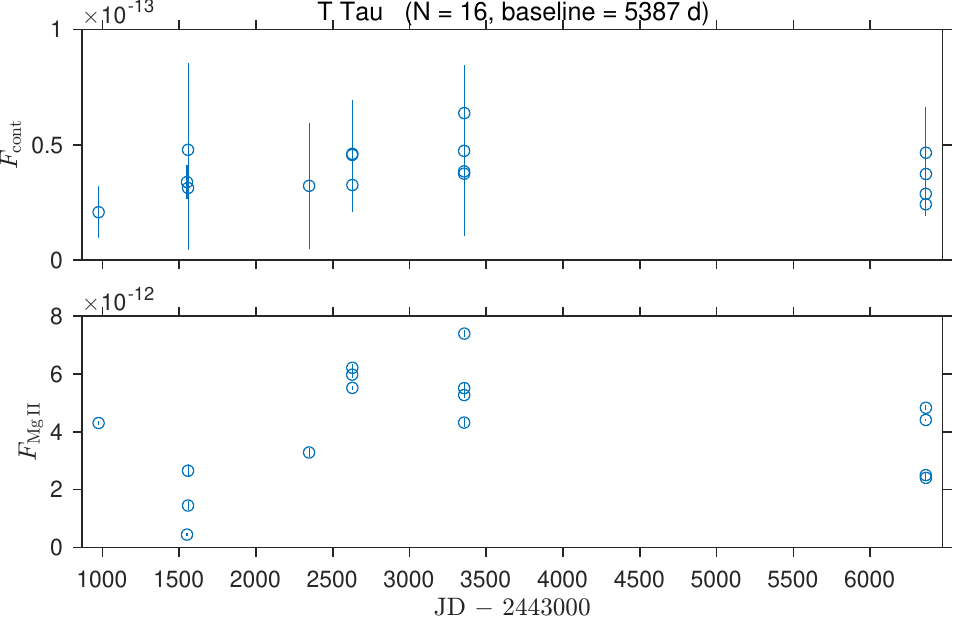} 
  \caption{Temporal sampling and variability of the near-UV continuum and Mg~II emission for the TTSs in this study. For each source, the upper panel shows the continuum flux interpolated beneath the Mg~II profile, in units of erg~s$^{-1}$~cm$^{-2}$~\AA$^{-1}$, and the lower panel shows the continuum-subtracted Mg~II line flux, in units of erg~s$^{-1}$~cm$^{-2}$. Error bars represent the uncertainties of the individual measurements.} 
  \label{fig:app_time1}
   \end{figure*}

\begin{figure*} 
   \centering 
   \includegraphics[width=0.45\textwidth]{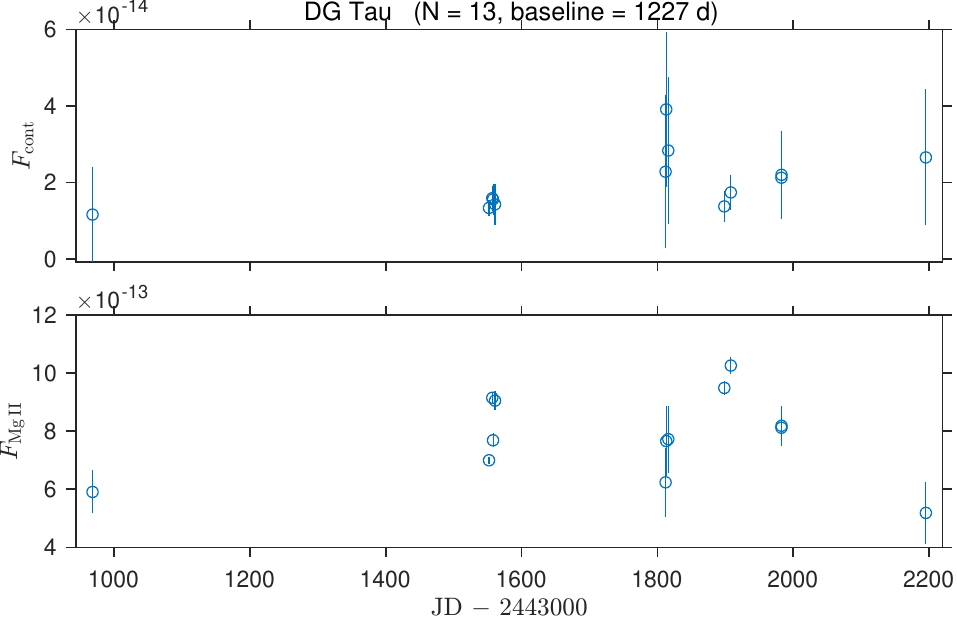} 
\vspace{0.25cm} 
   \includegraphics[width=0.45\textwidth]{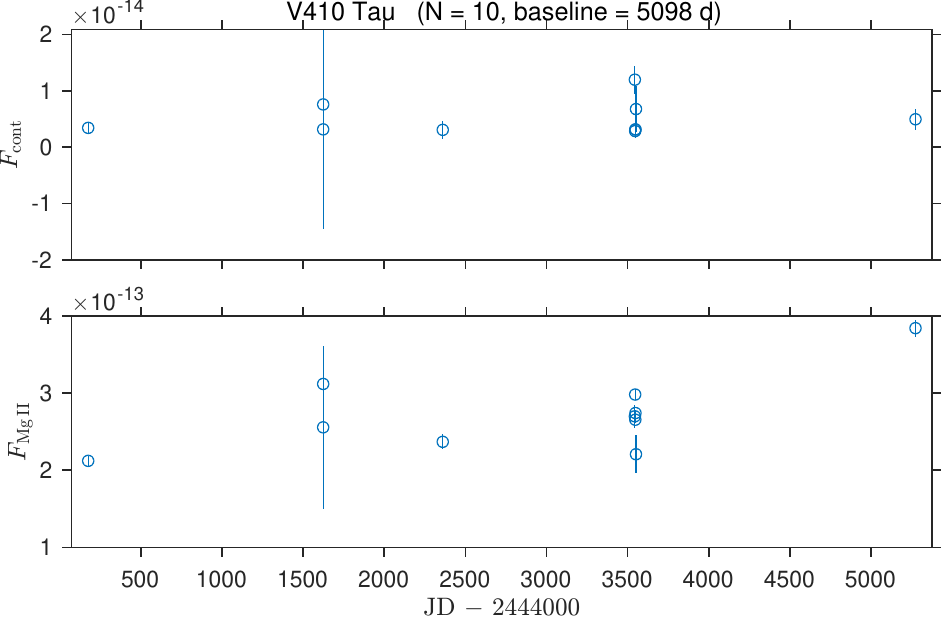} 
\vspace{0.25cm} 
   \includegraphics[width=0.45\textwidth]{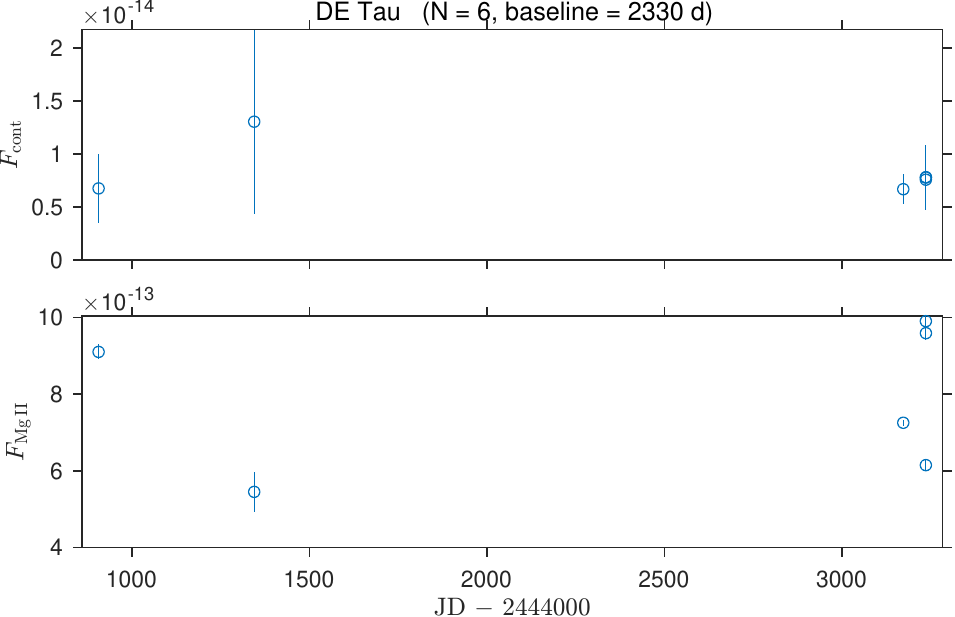} 
\vspace{0.25cm} 
   \includegraphics[width=0.45\textwidth]{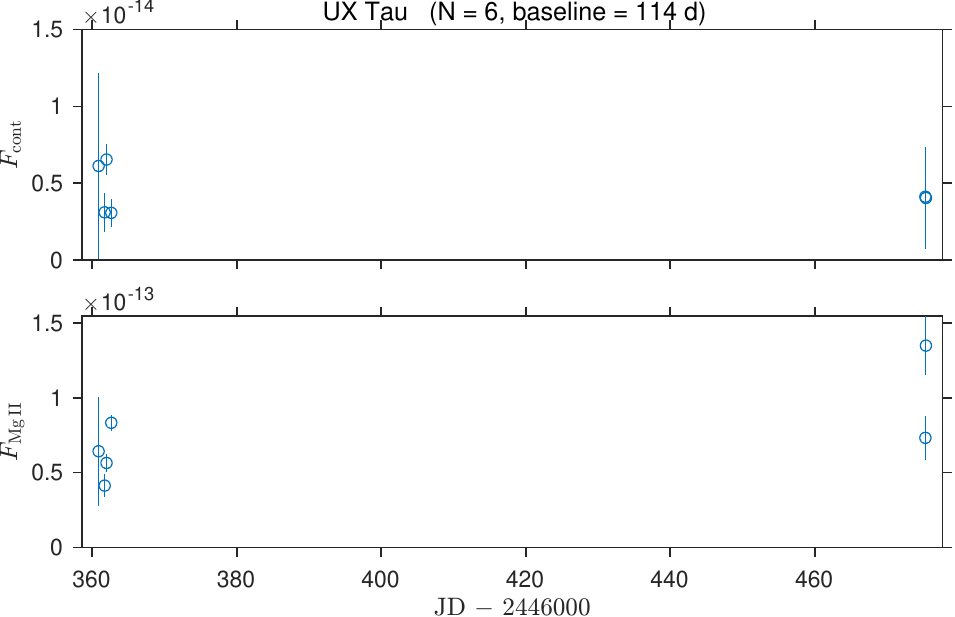} 
\vspace{0.25cm} 
   \includegraphics[width=0.45\textwidth]{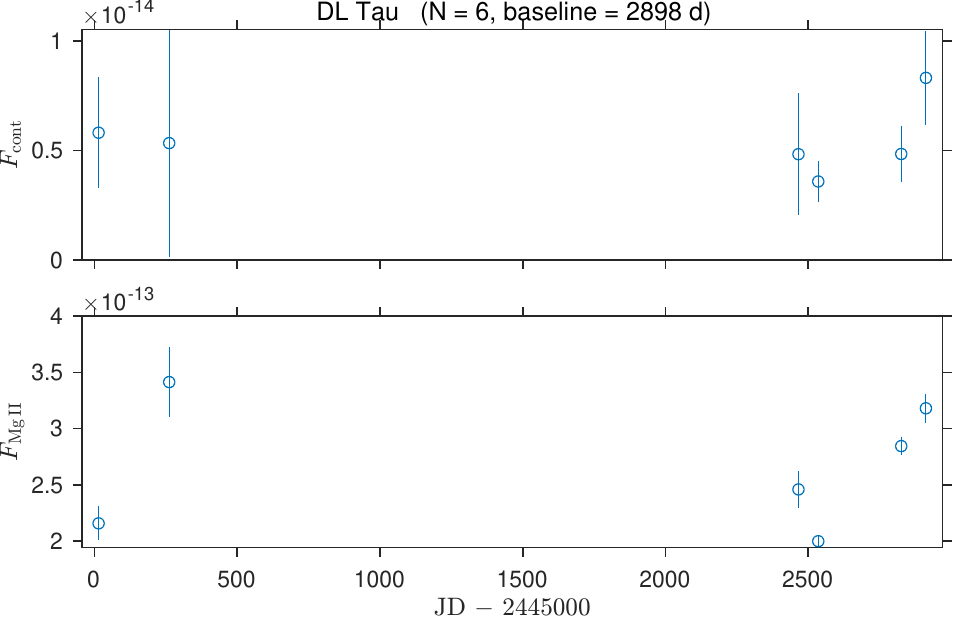} 
\vspace{0.25cm} 
   \includegraphics[width=0.45\textwidth]{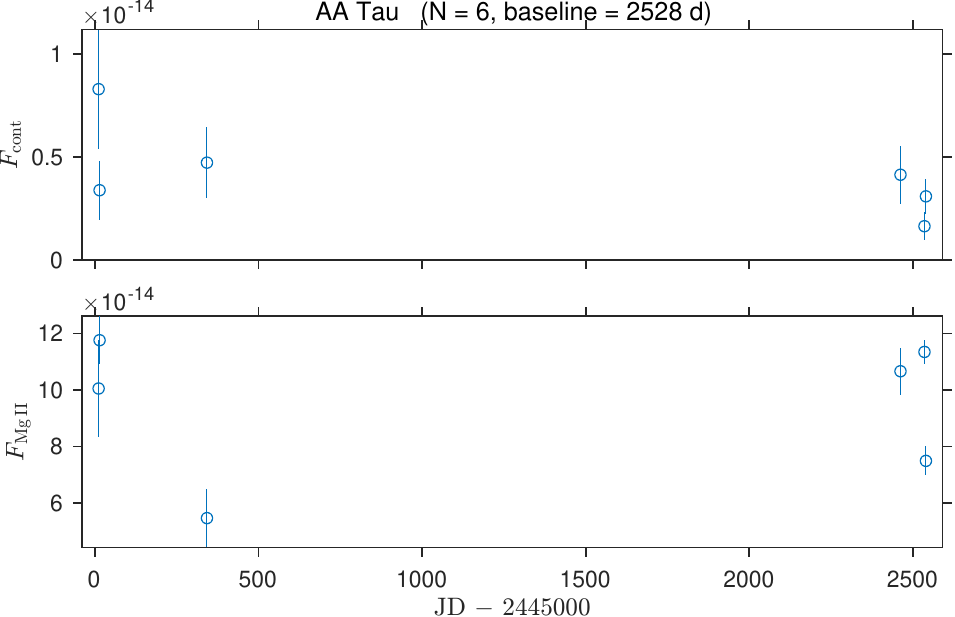} 
\vspace{0.25cm}
   \includegraphics[width=0.45\textwidth]{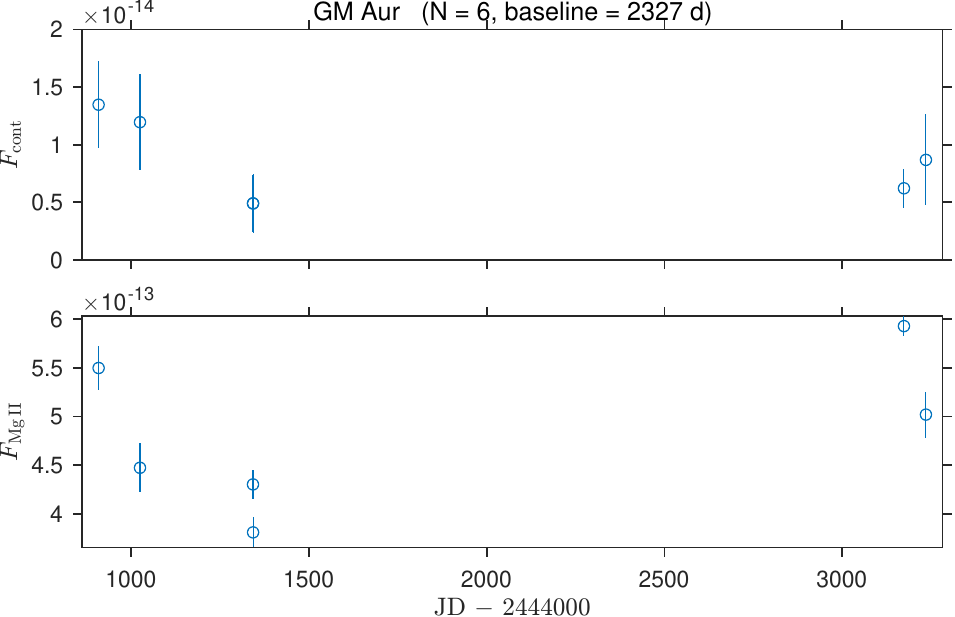} 
\vspace{0.25cm}
    \includegraphics[width=0.45\textwidth]{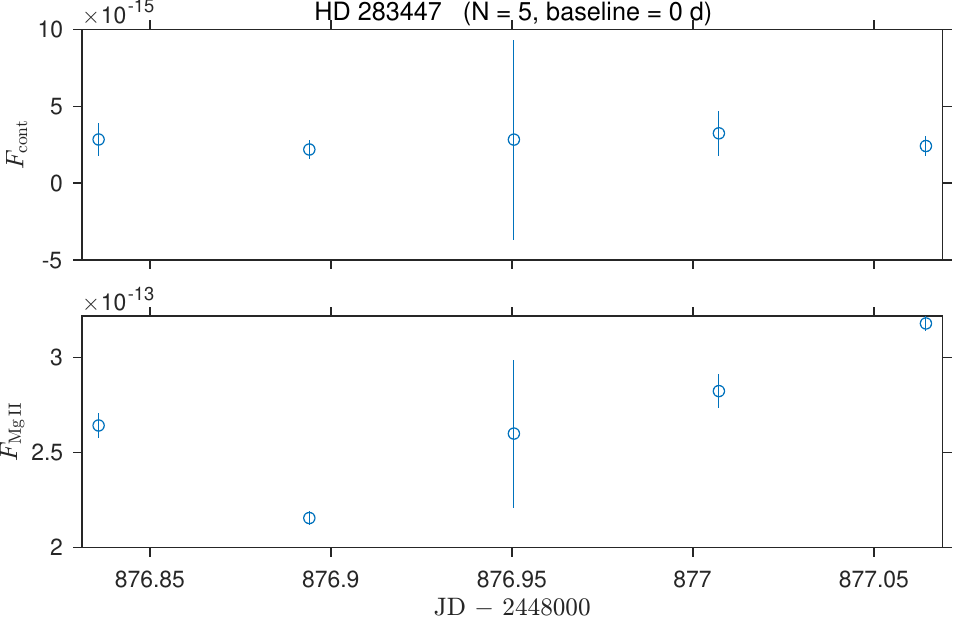} 
    
    \caption{Same as Fig.~\ref{fig:app_time1}, for the remaining sources in the sample.} 
    \label{fig:app_time2} 
    \end{figure*}

Figures~\ref{fig:app_time1}--\ref{fig:app_time2} show the temporal distribution of the IUE observations for the individual sources included in the analysis. The upper panel of each plot shows the continuum flux interpolated beneath the Mg~II profile, while the lower panel shows the continuum-subtracted Mg~II line flux. These figures provide a direct view of the strongly non-uniform temporal sampling of the IUE archive, including densely sampled observing intervals separated by long gaps.

   \end{appendix}

\end{document}